\documentstyle[12pt]{article}

\newcommand{\be}{\begin{equation}}
\newcommand{\ee}{\end{equation}}
\newcommand{\bea}{\begin{eqnarray}}
\newcommand{\eea}{\end{eqnarray}}
\newcommand{\bd}{\begin{displaymath}}
\newcommand{\ed}{\end{displaymath}}
\newcommand{\prim}{{\scriptscriptstyle \prime}}
\newcommand{\x}{\mbox{x}}
\begin{document}

\input FEYNMAN

\thispagestyle{empty}
\begin{flushright}
Roma1 n. 978 -1993
\end{flushright}
\begin{flushright}
ROM2F/93/38
\end{flushright}
\begin{flushright}
Revised version (May 1994)
\end{flushright}
\bigskip
\bigskip
\begin{center}
{\huge \bf Deep Inelastic Scattering}
\end{center}
\begin{center}
{\huge \bf in Improved Lattice QCD}
\end{center}
\begin{center}
{\Large \bf I. The first moment of structure functions}
\end{center}
\bigskip
\bigskip
\begin{center}
{\Large Stefano Capitani\ $^{1}$}
\end{center}
\medskip
\begin{center}
{\Large Giancarlo Rossi\ $^{2}$}
\end{center}
\bigskip
\bigskip
\bigskip
\bigskip

\ $^1$ Dipartimento di Fisica, Universit\`a degli Studi di Roma ``La
Sapienza'', and INFN, Sezione di Roma, P.le Aldo Moro 2, I-00185 Roma, Italy

\ $^2$ Dipartimento di Fisica, Universit\`a degli Studi di Roma
``Tor Vergata'', and INFN, Sezione di Roma 2,
Via della Ricerca Scientifica, I-00133 Roma, Italy

\bigskip
\bigskip

\begin{flushleft}
e-mail: \\
capitani@vaxrom.roma1.infn.it \\
rossig@vaxtov.roma2.infn.it
\end{flushleft}

\newpage
\thispagestyle{empty}

\begin{center}
{\large \bf ABSTRACT}
\end{center}

\newpage
\setcounter{page}{1}

\section{Introduction}

Deep Inelastic Scattering (DIS) experiments, which started at the end of the
sixties, shed light on the properties of strong interactions and on
the structure of hadrons (for a comprehensive review see \cite{part}).
The study of hard processes led to the formulation of the parton
model and to the hypothesis that quarks are the hadronic inner constituents
\cite{bjorken,feynman}. Further theoretical developments led to the
discovery of asymptotic freedom \cite{gross1,politzer} and eventually
to the formulation of Quantum Chromo-Dynamics \cite{qcd}. Today
DIS experiments are still a very important tool of investigation for short
distance physics phenomena and remain the best test of QCD as a theory of
strong interactions.

This paper is the first of two, focused on the computation of the
renormalization constants and mixing coefficients of the operators of
rank two and three that are related to the first two moments of
the DIS structure functions via the Wilson operator expansion of
the product of two weak or electromagnetic currents. The knowledge of
the hadronic matrix elements of the Wilson operators is necessary for the
theoretical evaluation of the moments of the $\x$-distribution of quarks and
gluons inside the hadrons. Our aim is to calculate to 1-loop renormalization
constants and mixing coefficients of the operators related to the first and
second moment in the framework of the nearest neighbor improved lattice QCD
\cite{She,rom1,rom2,rom3}.
These constants are needed to renormalize the lattice operators
and be able to extract physical hadronic matrix elements from numbers
obtained in Monte Carlo QCD simulations.

We present in this paper the general set up for these calculations and the
results for the operators related to the first moment, leaving the discussion
and the presentation of the results for the operator related to the second
moment to a forthcoming paper \cite{beccarini}. There are two main reasons
for the subdivision of the whole material in this way. One is that new
high-statistics Monte Carlo data for the first moment of DIS structure
functions, produced in simulations employing the improved lattice QCD action,
will soon be available.
The second reason, not unrelated with the previous one, is that, since
simulations concerning the second moment have been given lower priority, it has
not been yet decided what precise lattice expression for the corresponding rank
three operator will be finally taken. This is not immaterial from our point
of view: the perturbative calculation of the renormalization constants of the
rank three operators are barely within the reach of algebraic manipulation
programs such as Schoonschip or Form, because of the fantastically large
number of terms generated in the course of the calculation.

Renormalization constants and mixing coefficients of the rank two operators
have already been calculated some time ago using the standard Wilson action
with not completely consistent numerical results
\cite{kronfeld,Mart1,Mart2,Ross,pis1,pis2}.
In this work, beside checking previous calculations, we present the results
obtained for these constants by using the nearest neighbor improved QCD
action (also known as the ``clover-leaf'' action), according to the improvement
program started by Symanzik in ref.~\cite{Sym}.

The use of this action has been proven \cite{She,rom1,rom2,rom3} to remove
   from on-shell hadronic matrix elements all terms that in the continuum limit
are effectively of order $a$, $a$ being the lattice spacing. Recent Monte Carlo
simulations \cite{err1,err2} show indeed a substantial reduction
of the systematic errors related to the finiteness of the lattice spacing,
though at the price of a slightly higher theoretical and numerical
computational
effort.

Because of the great number of diagrams and of the complexity of the algebraic
expressions involved in the calculation even in the case of the rank two
operators, it has been considered necessary in this work to check all the
results obtained by hand by means of suitable algebraic manipulation programs.

The computation has also taken into account internal fermion loops,
hoping that the increasing power of dedicated computers
would soon allow for fully unquenched QCD simulations.

We have found discrepancies at the non-improved level, that is at the level
of the standard Wilson action, with some of the numbers already published in
the literature \cite{kronfeld,Mart1,Mart2,Ross}.
They have both numerical and analytic origin.
On the other hand, we are in complete agreement with the results of
refs.~\cite{pis1} and \cite{pis2} concerning the energy-momentum tensor.

This paper is organized as follows: in Sect.~2 we discuss the general
setting of the physical problem, and we immediately give the results for the
renormalization constants and mixing coefficients of the relevant operators in
a
``ready-to-use'' way. In Sect.~3 we present a brief introduction to the subject
of improved lattice perturbation theory. In Sect.~4 the properties of the
operators and their choice are discussed, in Sect.~5 we spell out all the
necessary renormalization conditions, in Sects.~6 and 7 some peculiar aspects
of the analytic and numerical calculations are explained, in Sect.~8 we present
the detailed results of our computation with a comparison with previous
results,
and in Sect.~9 we present some conclusions and an outlook of future lines
of investigations. Some of the more technical aspects of the calculations can
be found in the appendices A, B, C and D.

\pagebreak

\section{General setting and results}

\subsection{Moments of structure functions}

The DIS cross section is dominated by the field behavior in the light-like
region. The expansion of the product of, for instance,
two scalar field operators on the
light cone has the general expression \cite{Wilope}
\be
A(x) B(0) \sim \sum_{N,i} c_{N,i}(x^{2}) \, x^{\mu_1} \cdots x^{\mu_N}
O^{(N,i)}_{\mu_{1} \cdots \mu_{N}}(0) \label{eq:svcolu},
\ee
where the $O^{(N,i)}_{\mu_{1} \cdots \mu_{N}}$ are symmetric traceless tensors
with vanishing vacuum expectation value:
$ <0| O^{(N,i)}_{\mu_{1} \cdots \mu_{N}} |0> = 0$.
The singularity of the coefficient functions is, up to logarithms, given by
\be
c_{N,i}(x^{2}) \sim (x^{2})^{\frac{d_{O^{(N,i)}} - N - d_A - d_B}{2}}
\ee
and is governed by the twist \cite{twist},
\be
\tau = d_{O^{(N,i)}} - N ,
\ee
of the operator, that is, by the difference between its physical dimension
and the spin. The lowest twist operators are thus the dominant ones.

In QCD the lowest twist operators appearing in the light-cone expansion
of the product of two hadronic weak currents, relevant to DIS, are
\cite{gross,gross2,georgi,bardeen,kronfeld}
\bd
O^{qS}_{\mu_{1} \cdots \mu_{N}} = \frac{1}{2^N}
\overline{\psi} \: \gamma_{[ \mu_{1}} \!
\stackrel{\displaystyle \leftrightarrow}{D}_{\mu_{2}} \cdots
\stackrel{\displaystyle \leftrightarrow}{D}_{\mu_{N} ]} (1 \pm \gamma_5) \psi
\ed
\be
O^{qNS}_{\mu_{1} \cdots \mu_{N}} = \frac{1}{2^N}
\overline{\psi} \: \gamma_{[ \mu_{1}} \!
\stackrel{\displaystyle \leftrightarrow}{D}_{\mu_{2}} \cdots
\stackrel{\displaystyle \leftrightarrow}{D}_{\mu_{N} ]} (1 \pm \gamma_5)
\frac{\lambda^f}{2} \psi \label{eq:opsns}
\ee
\bd
O^{gS}_{\mu_{1} \cdots \mu_{N}} = \sum_{\rho} \mbox{Tr}
\left[ F_{[ \mu_{1} \rho}
\stackrel{\displaystyle \leftrightarrow}{D}_{\mu_{2}} \cdots
\stackrel{\displaystyle \leftrightarrow}{D}_{\mu_{N - 1}}
F_{\rho \mu_{N} ]} \right] ,
\ed
where the $\lambda^f$'s are flavor matrices. The operators (\ref{eq:opsns}) are
gauge invariant and all have $\tau = 2$; from now on we implicitly assume that
they are symmetrized with respect to all Lorentz indices. $S$ and $NS$
superscripts refer to Singlet and Non Singlet flavor structures.

In the unpolarized cross section the $\gamma_5$ contributions average to zero.
The other contributions have matrix elements of the form
\be
<p| O^{(N)}_{\mu_{1} \cdots \mu_{N}} |p> = A_N(\mu) p_{\mu_{1}} \cdots
p_{\mu_{N}}  + \mbox{trace terms} \label{eq:pppp},
\ee
where $\mu$ is the subtraction point.
They contain long distance non-perturbative physics, and are related to the
general expression of the moments of structure functions by the equations
\cite{christ}
\be
\int \! d\x \: \x^{N - 1} {\cal F}_{k}(q^{2},\x)= C_{N}(q^{2} / \mu^2)
A_{N} (\mu)
\label{eq:momtwist},
\ee
where we have put
\bea
{\cal F}_{1} & = 2 F_{1} \nonumber \\
{\cal F}_{2} & = \frac{\displaystyle F_{2}}{\displaystyle \x} \\
{\cal F}_{3} & = F_{3} \nonumber ,
\eea
with $F_1$, $F_2$ and $F_3$ the usual DIS structure functions as
defined in \cite{part}.
The Fourier transforms, $C_{N}$, of the Wilson coefficient
functions are determined \nolinebreak by the short distance perturbative
behavior of QCD and are formally given \nolinebreak by
\be
C_N(q^2)= (q^2)^N (\frac{\partial^2}{\partial q^2})^N
\int\! d^{4} x \: e^{\displaystyle i q x} \: c_N(x^{2}) \label{eq:four}.
\ee
When renormalization effects are taken into account, a dependence upon the
subtraction point $\mu$ appears in both sides of eq. (\ref{eq:four}).

The $C_N$'s are perturbatively calculable with renormalization group
techniques. In case of mixing among different operators
the Wilson coefficients satisfy a system of Callan-Symanzik equations
\be
\bigg( \mu \frac{d}{d \mu} + \beta (g) \frac{d}{d g} \bigg) C_{N,l} =
\sum_k (\gamma_N (g))_{l k} C_{N,k} \label{eq:calsi},
\ee
\nopagebreak
where the anomalous dimensions, $\gamma_N$, have been arranged in a matrix
\be
(\gamma_N)_{l k} = \mu \frac{d}{d \mu} \log (Z_N)_{l k}^{- 1} \label{eq:andim}.
\ee
Renormalization constants and mixing coefficients, $(Z_N)_{l k}$, are defined
by suitable sets of renormalization conditions,
allowing the construction of finite renormalized operators:
\be
\widehat{O}^{(N,l)}(\mu) = \sum_k (Z_N (\mu))_{l k} O^{(N,k)} .
\ee
Equations (\ref{eq:calsi}) can be decoupled by diagonalizing the anomalous
dimension matrix, so to have a new basis of operators that are multiplicatively
renormalizable.

\subsection{A summary of the results}

We have carried out, in the chiral limit, the 1-loop perturbative lattice
calculation of the renormalization constants and mixing coefficients of the
following rank two operators:
\be
O^q_{\mu \nu} = \frac{1}{4} \:\overline{\psi} \:\gamma_{[ \mu} \!
\stackrel{\displaystyle \leftrightarrow}{D}_{\nu ]} \psi \label{eq:o2} ,
\ee
which is related to the first moment of the $\x$-distribution of quarks,
and
\be
O^g_{\mu \nu} = \sum_{\rho} \mbox{Tr} \left[ F_{\mu \rho}
F_{\rho \nu} \right] \label{eq:og} ,
\ee
which is related to the first moment of the $\x$-distribution
of gluons\footnote{We are presently computing the renormalization
constant of the operator
\be
O^q_{\mu \nu \tau} = \frac{1}{8} \,\overline{\psi} \,\gamma_{[ \mu} \!
\stackrel{\displaystyle \leftrightarrow}{D}_{\nu}
\stackrel{\displaystyle \leftrightarrow}{D}_{\tau ]} \psi,
\ee
which is connected to the second moment of the $\x$-distribution of quarks
\cite{beccarini}.}.
The lattice expression of $F_{\mu \nu}$ is given by
\be
F_{n, \mu \nu} =
\frac{1}{8ig_0a^2} \sum_{\mu \nu = \pm} (U_{n, \mu \nu} - U^{+}_{n, \mu \nu}),
\ee
where $U_{n, \mu \nu}$ is the usual plaquette
\be
U_{n, \mu \nu} = U_{n,\mu} U_{n + \mu , \nu} U_{n + \nu , \mu}^{+}
U_{n, \nu}^{+}.
\ee
The operators (\ref{eq:o2}) and (\ref{eq:og}) are the first ones in the list
given above in eqs.~(\ref{eq:opsns}).

In the flavor Singlet sector, given the operators (\ref{eq:o2}) and
(\ref{eq:og}), the finite operators $\widehat{O}^1_{\mu \nu}$ and
$\widehat{O}^2_{\mu \nu}$, with well defined anomalous dimensions, that are
renormalized to their tree level values at $p^2 = \mu^2 \: (\mu a \ll 1)$, are
given to 1-loop in the full (unquenched) theory by the formulae (see
Sects.~5 and 8 for details)
\be
\widehat{O}^1_{\mu \nu} = \Bigg[ 1 - {\displaystyle\frac{\alpha_S}{4 \pi}}
\Big( N_c B_{gg} + N_f ( B^f_{gg} - B_{qg} ) \Big) \Bigg] O^g_{\mu \nu}
- \Bigg[ 1 + {\displaystyle\frac{\alpha_S}{4 \pi}}
C_F \Big( B_{gq} - B_{qq} \Big) \Bigg] O^q_{\mu \nu}
\ee
and
\bd
\widehat{O}^2_{\mu \nu} = \Bigg[ 1 - {\displaystyle\frac{\alpha_S}{4 \pi}}
\Big( \gamma_2 \log \mu a +  C_F B_{qq} +\frac{1}{4} N_f B_{gq}
\Big) \Bigg] O^q_{\mu \nu}
\ed
\be
+ {\displaystyle\frac{N_f}{4 C_F}} \Bigg[ 1 -
{\displaystyle\frac{\alpha_S}{4 \pi}}
\Big( \gamma_2 \log \mu a +  4 C_F B_{qg} + N_c B_{gg} + N_f B^f_{gg}
\Big) \Bigg] O^g_{\mu \nu} ,
\ee
\newline
where $N_c$  and $N_f$ are respectively the number of colors and the number
of flavors, and
\bea
\gamma_2 & = & \frac{16}{3} C_F + \frac{4}{3} N_f \nonumber \\
C_F & = & (N_c^2 - 1)/ 2N_c \nonumber \\
\alpha_S & = & g^2_0 / 4 \pi \label{eq:cfas}.
\eea

\begin{figure}
\begin{center}
\begin{tabular}{|c|r|c|r|}
\hline  \multicolumn{2}{|c}{Wilson} & \multicolumn{2}{|c|}{Improved} \\
\hline $B^W_{qq}$ & -3.165 & $B^I_{qq}$ & -15.816 \\
$B^W_{qg}$ & 0.019 & $B^I_{qg}$ & -1.041 \\
$B^W_{gq}$ & -5.817 & $B^I_{gq}$ & -4.044 \\
$(B^f_{gg})^W$ & -2.168 & $(B^f_{gg})^I$ & -19.425 \\
\hline
\multicolumn{4}{|c|}{$B_{gg} =$ -15.585} \\
\hline
\end{tabular}
\end{center}
\begin{center}
{\small {\bf Table 2.1} - Values of the constants $B$ on the lattice for
$r = 1$}
\end{center}
\bigskip
\bigskip
\bigskip
\begin{center}
\begin{tabular}{|c|r|r|}
\hline & sails & -4 \\
$B_{qq}$ & vertex & 8$/$9 \\
& self-energy & -1 \\
\cline{2-3} & \multicolumn{1}{|c|}{total} & - 37$/$9 \\
\hline \multicolumn{2}{|c}{$B_{qg}$} & \multicolumn{1}{|r|}{- 7$/$9} \\
\hline \multicolumn{2}{|c}{$B_{gq}$} & \multicolumn{1}{|r|}{- 23$/$9} \\
\hline \multicolumn{2}{|c}{$B_{gg}$} & \multicolumn{1}{|r|}{ 11$/$12} \\
\hline \multicolumn{2}{|c}{$B^f_{gg}$} & \multicolumn{1}{|r|}{- 23$/$18} \\
\hline
\end{tabular}
\end{center}
\begin{center}
{\small {\bf Table 2.2} - Values of the constants $B$ in $\overline{MS}$
regularization}
\end{center}
\end{figure}

The coefficients $B$ for the case of the standard Wilson action and of the
nearest neighbor improved action are reported (for $r = 1$) in Table 2.1.
They can be derived from the more detailed Tables presented in Sect.~8.
In Table 2.2 we report for completeness the values of the
constants $B$ in the continuum, computed in the $\overline{MS}$
renormalization scheme \cite{bardeen,Ross}.

In the quenched approximation one has to put $N_f = 0$ in the above formulae.
Numerically, in terms of $\beta \equiv 2 N_c / g_0^2$, for $N_c = 3$ one gets
\begin{itemize}
\item in the flavor Singlet sector

i) for the Wilson action:
\be
\widehat{O}^1_{\mu \nu} = \Bigg[ 1 - {\displaystyle\frac{9}{8 \pi^2 \beta}}
B_{gg} \Bigg] O^g_{\mu \nu} - \Bigg[ 1 + {\displaystyle\frac{1}{2 \pi^2 \beta}}
\Big( B^W_{gq} - B^W_{qq} \Big) \Bigg] O^q_{\mu \nu}
\ee
\bd
= \Bigg[ 1 + {\displaystyle\frac{1.776}{\beta}} \Bigg] O^g_{\mu \nu}
- \Bigg[ 1 - {\displaystyle\frac{0.134}{\beta}} \Bigg] O^q_{\mu \nu}
\ed
and
\be
\widehat{O}^2_{\mu \nu} = \Bigg[ 1 - {\displaystyle\frac{8}{3 \pi^2 \beta}}
\log \mu a  - {\displaystyle\frac{1}{2 \pi^2 \beta}} B^W_{qq} \Bigg]
O^q_{\mu \nu}
\ee
\bd
= \Bigg[ 1 - {\displaystyle\frac{0.270}{\beta}}
\log \mu a  + {\displaystyle\frac{0.160}{\beta}} \Bigg] O^q_{\mu \nu} ;
\ed

ii) for the nearest neighbor improved action:
\be
(\widehat{O}^1_{\mu \nu})^{IMPR} = \Bigg[ 1 - {\displaystyle\frac{9}{8 \pi^2
\beta}} B_{gg} \Bigg] O^g_{\mu \nu} - \Bigg[ 1 + {\displaystyle\frac{1}{2
\pi^2 \beta}} \Big( B^I_{gq} - B^I_{qq} \Big) \Bigg] (O^q_{\mu \nu})^{IMPR}
\ee
\bd
= \Bigg[ 1 + {\displaystyle\frac{1.776}{\beta}} \Bigg] O^g_{\mu \nu}
- \Bigg[ 1 + {\displaystyle\frac{0.596}{\beta}} \Bigg] (O^q_{\mu \nu})^{IMPR}
\ed
and
\be
(\widehat{O}^2_{\mu \nu})^{IMPR} = \Bigg[ 1 - {\displaystyle\frac{8}{3 \pi^2
\beta}} \log \mu a  - {\displaystyle\frac{1}{2 \pi^2 \beta}} B^I_{qq} \Bigg]
(O^q_{\mu \nu})^{IMPR}
\ee
\bd
= \Bigg[ 1 - {\displaystyle\frac{0.270}{\beta}}
\log \mu a  + {\displaystyle\frac{0.801}{\beta}} \Bigg] (O^q_{\mu \nu})^{IMPR}
,
\ed
where the explicit expression of $(O^q_{\mu \nu})^{IMPR}$ can be found in
eq.~(\ref{eq:o2impr}) of Sect.~6.

\item For the flavor Non Singlet operator
\be
O^{f q}_{\mu \nu} = \frac{1}{4}
\overline{\psi} \: \gamma_{[ \mu} \!
\stackrel{\displaystyle \leftrightarrow}{D}_{\nu ]}
\frac{\lambda^f}{2} \psi
\ee
there is no mixing between $O^q$ and $O^g$, and one gets
\be
\widehat{O}^{f q}_{\mu \nu} = \Bigg[ 1 - {\displaystyle\frac{8}{3 \pi^2 \beta}}
\log \mu a  - {\displaystyle\frac{1}{2 \pi^2 \beta}} B_{qq} \Bigg]
O^{f q}_{\mu \nu} ,
\ee
with obvious adjustments in notations in going from the Wilson to the
improved case and with $B_{qq}$ given in Table 2.1.
\end{itemize}

\section{Improved lattice QCD}

Lattice QCD represents today the only viable way of evaluating from first
principles the matrix elements (\ref{eq:pppp}), needed for the computation of
$A_N$ and, hence, of the moments of the structure functions.

In this section we wish to briefly describe the improvement program of lattice
QCD, as discussed in refs.~\cite{Sym,Lus,She,rom1,rom2,rom3}.

The QCD Wilson action for one flavor $f$ is, on a euclidean lattice
\cite{Wil},
\bd
S^{f}_{LATT} = a^4 \sum_{n\ } \Bigg[  - \frac{1}{2 a} \sum_{\mu} \Big[
\overline{\psi}_{n} ( r - \gamma_{\mu} ) U_{n,\mu} \psi_{n + \mu}
\ed
\bd
+ \overline{\psi}_{n + \mu} ( r + \gamma_{\mu} ) U_{n,\mu}^{+} \psi_{n}
\Big] + \overline{\psi}_{n} \left( m_{f} + \frac{4 r}{a} \right) \psi_{n}
\Bigg]
\ed
\be
- {\displaystyle\frac{1}{g_0^2}} \sum_{n,\mu \nu}
\Bigg[ \mbox{Tr} \left[ U_{n,\mu} U_{n + \mu , \nu} U_{n + \nu , \mu}^{+}
U_{n, \nu}^{+} \right] - N_c \Bigg] \label{eq:wil} .
\ee
The gauge field $A_{n,\mu}$ is introduced through the definition
\be
U_{n,\mu} = e^{\displaystyle i g_0 a t^{A} A_{n,\mu}^{A}} \ \ \ \
(A = 1 , \ldots , N_c^{2} -1 ).
\ee
Fermion fields are site variables, while the $U_{n,\mu}$'s are link variables.
$U_{n,\mu}$ is associated to the link connecting the site $n$ with the
site $n + \mu$. From this action one can derive the Feynman rules for lattice
perturbation theory. Conventionally $A_{n,\mu}$ will be taken to sit at the
point $n + \mu / 2$, the middle point of the link ($n,n+\mu$) (see Fig.~1).
This choice greatly simplifies Feynman rules.

\begin{figure}
\begin{center}
\begin{picture}(20000,15000)
\drawline\fermion[\E\REG](5000,2500)[10000]
\put(10000,2500){\circle*{300}}
\put(10000,500){$U_{n,\mu}$}
\drawarrow[\LDIR\ATTIP](12500,2500)
\THICKLINES
\drawline\fermion[\E\REG](10000,2500)[2500]
\thinlines
\put(15000,2500){\circle*{500}}
\put(17000,500){$\psi_{n+\mu}$}
\drawline\fermion[\N\REG](15000,2500)[10000]
\put(15000,7500){\circle*{300}}
\put(17000,7500){$U_{n+\mu,\nu}$}
\drawarrow[\LDIR\ATTIP](15000,10000)
\THICKLINES
\drawline\fermion[\N\REG](15000,7500)[2500]
\thinlines
\put(15000,12500){\circle*{500}}
\put(17000,14500){$\psi_{n+\mu+\nu}$}
\drawline\fermion[\W\REG](15000,12500)[10000]
\put(10000,12500){\circle*{300}}
\put(9000,14500){$U^+_{n+\nu,\mu}$}
\drawarrow[\LDIR\ATTIP](7500,12500)
\THICKLINES
\drawline\fermion[\W\REG](10000,12500)[2500]
\thinlines
\put(5000,12500){\circle*{500}}
\put(2000,14500){$\psi_{n + \nu}$}
\drawline\fermion[\S\REG](5000,12500)[10000]
\put(5000,7500){\circle*{300}}
\put(1000,7500){$U^+_{n,\nu}$}
\drawarrow[\LDIR\ATTIP](5000,5000)
\THICKLINES
\drawline\fermion[\S\REG](5000,7500)[2500]
\thinlines
\put(5000,2500){\circle*{500}}
\put(2000,500){$\psi_n$}
\end{picture}
\end{center}
\begin{center}
{\small {\bf Fig. 1} - The $\mu\nu$-plaquette in lattice QCD}
\end{center}
\end{figure}

The computations of physical quantities with Monte Carlo simulations are
affected by statistical and systematic errors. The one which we are mainly
concerned with here is the systematic error due to the finiteness of the
lattice spacing. In fact for a renormalized finite lattice operator,
$\widehat{\cal O}_{L}$, we have the formal expansion
\be
\left< p \left| \widehat{\cal O}_{L} \right| p^{\prim} \right>_{Monte\ Carlo}=
a^{d} \left[ \left< p \left| \widehat{\cal O} \right| p^{\prim} \right>_{phys.}
+ O(a) \right] \label{eq:a} ,
\ee
where $\left< p \left| \widehat{\cal O} \right| p^{\prim} \right>_{phys.}$
is the physical
matrix element we want to extract from Monte Carlo data and $d$ is its physical
dimension. We see from equation (\ref{eq:a}) that the possibility of estimating
$\left< p \left| \widehat{\cal O} \right| p^{\prim} \right>_{phys.}$
      from Monte Carlo simulations depends on the magnitude of $O(a)$ terms,
which is generally between 20 and 30 percent. The effect of improvement
is the removal of all corrections that in the continuum limit ($g_0^2 \sim
1/ \log a$) are effectively of order $a$, that is to say, of all terms that
in perturbation theory are of the type
\be
a (g_0^2)^n (\log a)^n \sim \mbox{``}a\mbox{''} \label{eq:limef}.
\ee
This result is achieved through the addition of ``irrelevant'' interaction
terms to the Wilson action and by adjusting (``improving'') the lattice
expression of fermion operators, so as to cancel unwanted contributions in
their matrix elements \cite{Sym},
thus making faster the recovery of the continuum properties in the limit
$a \rightarrow 0$. It is estimated that with this method one can achieve a
reduction of the systematic error due to the finiteness of the lattice spacing
down to 5 - 10 percent \cite{err1,err2}.

The first proposal of improvement of the fermionic part of the QCD action
presented, however, the drawback of having next-to-nearest neighbor
interactions
\cite{sec1,sec2,sec3}. This feature is a problem in Monte Carlo
simulations, given the calculational effort required for the numerical
inversion of the fermion propagator (to date the only simulation carried out
with the next-to-nearest neighbor improved action is that of
ref.~\cite{secsim}). A great step forward was made
by L\"uscher and Weisz, who introduced the notion of on-shell improvement
\cite{Lus}: on-shell improvement is a weaker requirement than full
improvement and easier to implement. Immediately after, Sheikholeslami and
Wohlert \cite{She} proposed the nearest
neighbor fermion action employed today and started the study of its properties.
In ref.~\cite{rom1} the relation between the Sheikholeslami-Wohlert action and
the next-to-nearest action was elucidated and it
was shown that the Sheikholeslami-Wohlert action leads to on-shell improvement,
with the consequence that all spectral quantities are free from terms
that in the continuum limit are effectively of order ``$a$''
(eq.~(\ref{eq:limef})), provided that hadronic operators are at the same time
appropriately ``improved''.

The developments of these researches on on-shell improvement led to a
practical recipe for its application to QCD
\cite{rom1,rom2,rom3}. This consists in modifying the standard Wilson action
by adding the nearest-neighbor interaction term
\be
\Delta S^{f}_{I} = - i g_0 a^{4} \sum_{n,\mu \nu} \frac{r}{4 a} \:
\overline{\psi}_{n}
\sigma_{\mu \nu} F_{n, \mu \nu} \psi_{n} \label{eq:impr}.
\ee
Here $F_{n, \mu \nu}$ is not the naive lattice ``plaquette''
\be
P_{n, \mu \nu} = \frac{1}{2ig_0a^2} (U_{n, \mu \nu} - U^{+}_{n, \mu \nu})
\label{eq:effeno}
\ee
\bd
U_{n, \mu \nu} = U_{n,\mu} U_{n + \mu , \nu} U_{n + \nu , \mu}^{+}
U_{n, \nu}^{+},
\ed
but rather the average of the four plaquettes lying in the plane $\mu \nu$,
stemming from the point $n$:
\be
F_{n, \mu \nu} = \frac{1}{4} \sum_{\mu \nu = \pm} P_{n, \mu \nu} =
\frac{1}{8ig_0a^2} \sum_{\mu \nu = \pm} (U_{n, \mu \nu} - U^{+}_{n, \mu \nu}).
\label{eq:effesi}
\ee
Because of the form of $F_{n, \mu \nu}$, the Sheikholeslami-Wohlert action is
often referred to as the ``clover-leaf'' action.

The term (\ref{eq:impr}), because of the antisymmetry of
$\sigma_{\mu \nu}$ matrices,
does not alter the quark-gluon interactions with
an even number of gluons. The interactions with an odd number of gluons are
on the contrary modified. For 1-loop calculations this results in the
appearance of the new vertex
\be
(V^I)^{bc}_{\rho} (k,k') = -g_0 \frac{r}{2} (t^A)_{bc}
\cos \frac{a(k - k')_{\rho}}{2} \sum_{\lambda}
\sigma_{\rho \lambda} \sin a(k - k')_{\lambda} \label{eq:newimpr},
\ee
where $k$ and $k'$ are the momenta of the incoming and of the outgoing
fermions,
and $\rho$ is the Lorentz index carried by the gluon, to be added to the
standard Wilson vertex
\be
(V)^{bc}_{\rho} (k,k') = -g_0 (t^A)_{bc} \left[ r \sin \frac{a(k + k')_{\rho}}
{2} + i \gamma_{\rho} \cos \frac{a(k + k')_{\rho}}{2} \right] .
\ee

Besides this modification of the action, in the calculation of an $n$-point
fermion Green function, one has to perform on each fermion field the rotation
(see appendix A for the definition of the lattice covariant derivatives)
\bd
\psi \longrightarrow \left( 1 - \frac{a r}{2} \stackrel{\displaystyle
\rightarrow }{\not\!\!{D}} \right) \psi
\ed
\be
\overline{\psi} \longrightarrow \overline{\psi} \left( 1 + \frac{a r}{2}
\stackrel{\displaystyle \leftarrow }{\not\!\!{D}} \right) \label{eq:rotation}.
\ee

As we said, it was proven in ref.~\cite{rom1} that this recipe allows to get
rid in on-shell matrix elements of all terms that in perturbation theory are of
the form (\ref{eq:limef}). The removal of these leading log terms
leaves us with next-to-leading terms of the kind
\be
a (g_0^2)^n \log^{n - 1} a ,
\ee
which are effectively of order ``$a / \log a$''. The improvement recipe
thus lowers the difference between continuum and lattice from
(\ref{eq:a})
to
\be
\left< p \left| \widehat{\cal O}_{L} \right| p^{\prim}
\right>^{IMPR.}_{Monte\ Carlo}=
a^{d} \left[ \left< p \left| \widehat{\cal O} \right| p^{\prim} \right>_{phys.}
+ O(a/\log a)
\right] .
\ee

Getting rid of also the next-to-leading terms would require a bunch of
\linebreak 1-loop
computations. They are necessary to adjust to next order the coefficient
in front
of eq.~(\ref{eq:impr}) and the form of the transformation (\ref{eq:rotation}).
A crucial step forward in this direction has been carried out in
ref.~\cite{naik} where the $g_0^2$ correction to the factor in front of
eq.~(\ref{eq:impr}) has been computed. This is enough to improve to
next-to-leading log's the values of the hadron masses measured
in Monte Carlo simulations \cite{khadra,khadra2}.

\section{The operators}

We have computed, in the chiral limit, the four forward matrix elements
\be
\left( \begin{array}{cc}
<q| O^q_{\mu \nu} |q> & <g,\sigma| O^q_{\mu \nu} |g,\sigma> \\
<q| O^g_{\mu \nu} |q> & <g,\sigma| O^g_{\mu \nu} |g,\sigma>
\end{array} \right) \label{eq:matrix},
\ee
where $|q>$ is a one-quark state of momentum $p$ and vanishing (renormalized)
mass and $|g,\sigma>$ is a one-gluon state of momentum $p$ and polarization
$\sigma$.

It should be
noticed that beyond tree level there will be in general a mixing in the
flavor Singlet sector between the quark operator $O^q$ and the gluon operator
$O^g$. However, in the ``quenched'' approximation ($N_f =$ number of flavors
$= 0$) one has $<g,\sigma| O^q_{\mu \nu} |g,\sigma> = 0$. In this
case the matrix (\ref{eq:matrix}) becomes triangular and the operator $O^q$
will not mix anymore with $O^g$ (see eqs.~(\ref{eq:eigva}), (\ref{eq:eigve1})
and (\ref{eq:eigve2}) below).

To write the operators (\ref{eq:o2}) and (\ref{eq:og}) in a
euclidean lattice, we need to take into account the changes in the
symmetry properties of the Lagrangian \cite{mix1,mix2} due to the the
breaking of
the euclidean Lorentz group O(4) down to the discrete hypercubic group H(4):
\be
O(4) \longrightarrow H(4)  \label{eq:breaking}
\ee
\bd
(\mbox{continuum} \rightarrow \mbox{cubic lattice}) .
\ed

The hypercubic group H(4) is the group of the discrete rotations of the lattice
onto itself. It is a finite subgroup of the ortogonal four dimensional group
O(4), consisting of 192 elements. H(4) admits 13 irreducible representations.

The breaking (\ref{eq:breaking}) implies that the operators belonging to
irreducible representations of O(4) may transform in a reducible way under
H(4),
thus inducing a mixing under renormalization among operators
belonging to different irreducible O(4) representations.
This mixing adds to the mixing between (\ref{eq:o2}) and (\ref{eq:og}), and
causes further complications in the
computation, even if the new operators have the same dimensions as the
original one. If they have lower dimensions the coefficient of the mixing will
inevitably contain negative powers of $a$, giving rise to power divergences:
they can only be subtracted from the original operator in a
non-perturbative way \cite{bochicchio}.

To avoid all these unwanted mixings we need to choose particular values of the
$\mu\nu$-indices in eqs.~(\ref{eq:o2}) and (\ref{eq:og}),
together with an appropriate lattice expression for the gauge field strength,
$F_{\mu \nu}$, in eq.~(\ref{eq:og}).

First, we note that actually the quark operator has the explicit form
\be
O^q_{\mu \nu} = \frac{1}{4} \:\overline{\psi} \:\gamma_{[ \mu} \!
\stackrel{\displaystyle \leftrightarrow}{D}_{\nu ]} \psi  -
\delta_{\mu \nu} \cdot \mbox{tr} \label{eq:trs},
\ee
and the like for the gluon case (\ref{eq:og}). The subtraction of
trace terms, symbolically indicated by ``tr'' in eq.~(\ref{eq:trs}),
reduces eq.~(\ref{eq:pppp}) to the simpler form
\be
<p| O^{(N)}_{\mu_{1} \cdots \mu_{N}} |p> = A_{N} \,  p_{\mu_{1}}
\cdots p_{\mu_{N}}  .
\ee
The operator $O^q_{\mu \nu}$ with no trace subtraction is
the sum of two irreducible representations of H(4): a four-dimensional one,
corresponding to $\mu = \nu$, and a six-dimensional one, corresponding to
$\mu \neq \nu$. With the subtraction of the trace the singlet is removed from
the four-dimensional representation, which therefore becomes three-dimensional
and cannot mix any more with the dangerous lower dimensional singlet operator
$\overline{\psi} \:\psi$ with a linearly divergent coefficient.
However, large errors would result when the subtraction of the trace
contribution is numerically performed on Monte Carlo data.
The other option is therefore to choose $\mu\neq\nu$. As we said above, this
leads to an irreducible H(4) representation, so there will be no mixing with
other operators.
With this choice it will be necessary to give the hadron a momentum
different from zero in one of the two $\mu$ or $\nu$ directions. This
causes a somewhat stronger sensitivity of Monte Carlo data to the granularity
of the lattice.

In actual Monte Carlo simulations
\cite{Mart1,Mart2} the choice $\mu\neq\nu$ has always been made, as the
magnitude of the systematic errors associated with it is smaller than the one
coming from the numerical subtraction of the trace contribution. In this paper
also we will work with $\mu\neq\nu$. This, by the way, induces remarkable
simplifications in the calculations.

We further restricted ourselves to the chiral case, that is to the case in
which the (renormalized) quark mass is zero. This choice corresponds to the
limiting situation of vanishing pion mass, to which Monte Carlo data are always
extrapolated.

As for the gluon operator, the same lattice approximation of the gluon
field strength that is chosen in (\ref{eq:effesi}) turns out to
avoid mixing of $O^g$ with undesired operators \cite{mix1,mix2}. In fact
the operator (\ref{eq:effeno}) transforms like a 24-dimen\-sional
reducible representation and the product
$P_{\mu \rho} \cdot P_{\rho \nu}$, with $\mu\neq\nu$, transforms like a
192-dimensional reducible representation.
In particular $P_{\mu \rho} \cdot P_{\rho \nu}$ mixes with the operator
$\overline{\psi} \:\psi$ with a power divergent coefficient. On the contrary,
with the definition (\ref{eq:effesi}), which maximizes the symmetry of the
expression of the field strength on the lattice, $F_{n, \mu \nu}$ will
transform like the direct sum of two three-dimensional irreducible
representations corresponding to the two combinations $E + B$ and $E - B$.
Also in this case we choose $\mu \neq \nu$ to avoid the need for trace
subtraction.

\section{Renormalization conditions}

The renormalization conditions connect the bare lattice operators on the
lattice to finite operators renormalized at a scale $\mu$:
\be
\widehat{O}^l(\mu) = Z_{lk}(\mu a) O^k(a) \label{eq:cosrin}.
\ee
The constants $Z_{lk}$ are fixed in perturbation theory by the same
renormalization conditions used in the continuum.
As we have discussed in the previous section, in the flavor Singlet case there
is a mixing between the quark and gluon operators (\ref{eq:o2}) and
(\ref{eq:og}) that have the same conserved quantum numbers.
However, with the choice $\mu \neq \nu$ and the
definition (\ref{eq:effesi}) of the gauge field strength we avoid mixing with
other operators of possibly lower dimensions, and therefore the need for
subtraction of power divergences.

We thus write:
\bd
\widehat{O}^q = Z_{qq} O^q + Z_{qg} O^g
\ed
\be
\widehat{O}^g = Z_{gq} O^q + Z_{gg} O^g \label{eq:ref1},
\ee
where the $Z$'s are determined by imposing the renormalization conditions:
\bea
<q| \widehat{O}^q (\mu) |q> & = & <q| O^q (a) |q> |^{tree}_{p^2 = \mu^2}
\nonumber \\
<g,\sigma| \widehat{O}^q (\mu) |g,\sigma>
& = & <g,\sigma| O^q (a) |g,\sigma> |^{tree}_{p^2 = \mu^2} = 0  \nonumber \\
<q| \widehat{O}^g (\mu) |q>
& = &  <q| O^g (a) |q> |^{tree}_{p^2 = \mu^2} = 0 \nonumber \\
<g,\sigma| \widehat{O}^g (\mu) |g,\sigma>
& = & <g,\sigma| O^g (a) |g,\sigma> |^{tree}_{p^2 = \mu^2} \label{eq:ref2}.
\eea

In the perturbative computation of the two members of eqs.~(\ref{eq:ref2}) we
have in our hands the choice of the polarization index $\sigma$ of the external
gluon (from which the $Z$'s obviously do not depend). To simplify our
successive calculations we decided to take $\sigma \neq \mu,\nu$.

In the tree approximation the amputated non vanishing matrix elements of the
operators $O^q_{\mu \nu}$ and $O^g_{\mu \nu}$ are
given (for $\mu \neq \nu$, $\mu \neq \sigma$, $\nu \neq \sigma$) by
\bea
<q| O^q_{\mu \nu} (a) |q> |^{tree}_{amp} & = & \frac{1}{2} \Bigg[ \frac{i}{2}
\gamma_{\mu} p_{\nu} + (\mu \rightarrow \nu ) \Bigg] \nonumber \\
<g,\sigma| O^g_{\mu \nu} (a) |g,\sigma> |^{tree}_{amp} & = &
- p_{\mu} p_{\nu} \label{eq:ref3}.
\eea
  From eqs.~(\ref{eq:ref1}) and (\ref{eq:ref2})  the renormalization constants
and the mixing coefficients of the rank two operators can be written for
$\mu a \ll 1$ in the form \cite{gross,gross2}
\bea
Z_{qq} (\mu a) & = & 1 - {\displaystyle\frac{\alpha_S}{4 \pi}} C_F
\left( {\displaystyle\frac{16}{3}} \log \mu a + B_{qq} \right)
\nonumber \\
Z_{qg} (\mu a) & = & - {\displaystyle\frac{\alpha_S}{4 \pi}} N_f
\left( {\displaystyle\frac{4}{3}} \log \mu a + B_{qg} \right)
\nonumber \\
Z_{gq} (\mu a) & = & - {\displaystyle\frac{\alpha_S}{4 \pi}} C_F
\left( {\displaystyle\frac{16}{3}} \log \mu a + B_{gq} \right)
\nonumber \\
Z_{gg} (\mu a) & = & 1 - {\displaystyle\frac{\alpha_S}{4 \pi}} \left[ N_f
\left( {\displaystyle\frac{4}{3}} \log \mu a +
B^f_{gg} \right) + N_c B_{gg} \right] \label{eq:zeta} ,
\eea
where $C_F$ and $\alpha_S$ have been defined in eqs.~(\ref{eq:cfas}).

The coefficients of the logarithms are the so called anomalous dimensions.
They are physical quantities and obviously their values are not affected by
the improvement procedure.
The difference between the use of improved and non-improved action
lies only in the finite constants $B$. Our results for the $B$'s are reported
in Sect.~8.

If we want operators with well defined anomalous dimensions, i.e. operators
that
are multiplicatively renormalizable, we must diagonalize the matrix
(\ref{eq:andim}) of anomalous dimensions. Its determinant, as it can be seen
     from (\ref{eq:zeta}), is zero. The two eigenvalues are
\bea
\gamma_1 & = & 0 \nonumber \\
\gamma_2 & = & \frac{16}{3} C_F + \frac{4}{3} N_f \label{eq:eigva},
\eea
and the corresponding eigenvectors
\be
\widehat{O}^1_{\mu \nu} = \Bigg[ 1 - {\displaystyle\frac{\alpha_S}{4 \pi}}
\Big( N_c B_{gg} + N_f ( B^f_{gg} - B_{qg} ) \Big) \Bigg] O^g_{\mu \nu}
- \Bigg[ 1 + {\displaystyle\frac{\alpha_S}{4 \pi}}
C_F \Big( B_{gq} - B_{qq} \Big) \Bigg] O^q_{\mu \nu}  \label{eq:eigve1}
\ee
and
\bd
\widehat{O}^2_{\mu \nu}  = \Bigg[ 1 - {\displaystyle\frac{\alpha_S}{4 \pi}}
\Big( \gamma_2 \log \mu a +  C_F B_{qq} +\frac{1}{4} N_f B_{gq}
\Big) \Bigg] O^q_{\mu \nu}
\ed
\be
+ {\displaystyle\frac{N_f}{4 C_F}} \Bigg[ 1 -
{\displaystyle\frac{\alpha_S}{4 \pi}}
\Big( \gamma_2 \log \mu a +  4 C_F B_{qg} + N_c B_{gg} + N_f B^f_{gg}
\Big) \Bigg] O^g_{\mu \nu}  \label{eq:eigve2}.
\ee

The vanishing of one eigenvalue ($\gamma_1$) of the anomalous dimension matrix
means that there exists a conserved operator ($\widehat{O}^1$). Physically
$\widehat{O}^1$ corresponds to the full energy-momentum tensor of the system.
Notice also that in the quenched approximation ($N_f = 0$) the operator
$\widehat{O}^2$ is simply proportional to $O^q$, as no mixing is possible with
$O^g$ in absence of quark loops.

\section{Details of the computation}

The improvement prescription discussed in Sect.~3 requires that in the
fermion operators the spinor fields have to be rotated according to
eqs.~(\ref{eq:rotation}). This means that the improved
operators are much more complicated than the non-improved ones.
For instance the rank two quark operator explicitly becomes
\bd
(O^q_{\mu \nu})^{IMPR} =
\frac{1}{4} \left[ \overline{\psi} \,\gamma_{\mu} \!
\stackrel{\displaystyle \rightarrow}{D}_{\nu} \psi
- (\overline{\psi} \stackrel{\displaystyle \leftarrow}{D}_{\nu})
\,\gamma_{\mu} \psi \right]
\ed
\bd
- \frac{a r}{8} \left[\overline{\psi} \,\gamma_{\mu} \!
\stackrel{\displaystyle \rightarrow}{D}_{\nu} \stackrel{\displaystyle
\rightarrow }{\not\!\!{D}} \psi -(\overline{\psi} \stackrel{\displaystyle
\leftarrow }{\not\!\!{D}}) \gamma_{\mu} \!
\stackrel{\displaystyle \rightarrow}{D}_{\nu} \psi \right.
- \left.(\overline{\psi} \stackrel{\displaystyle \leftarrow}{D}_{\nu})
\,\gamma_{\mu} \stackrel{\displaystyle\rightarrow }{\not\!\!{D}}  \psi
+ (\overline{\psi} \stackrel{\displaystyle
\leftarrow }{\not\!\!{D}} \stackrel{\displaystyle \leftarrow}{D}_{\nu})
\,\gamma_{\mu} \psi  \right]
\ed
\be
- \frac{a^2 r^2}{16} \left[ (\overline{\psi} \stackrel{\displaystyle
\leftarrow }{\not\!\!{D}}) \,\gamma_{\mu} \stackrel{\displaystyle \rightarrow}
{D}_{\nu} \stackrel{\displaystyle
\rightarrow }{\not\!\!{D}} \psi
- (\overline{\psi} \stackrel{\displaystyle
\leftarrow }{\not\!\!{D}} \stackrel{\displaystyle \leftarrow}{D}_{\nu})
\,\gamma_{\mu} \stackrel{\displaystyle
\rightarrow }{\not\!\!{D}} \psi \right] \label{eq:o2impr}.
\ee

It should be noted that, although we are dealing with order $a$ improvement, in
calculating the $Z$'s of eqs.~(\ref{eq:ref1}) we have to take into account also
the contributions formally of order $a^2$ that arise when the rotations
on $\psi$ e $\overline{\psi}$ are both performed at the same time
\cite{borrelli}.
These terms in fact do contribute to the constants $B$ because $1/a^2$
divergences present in 1-loop lattice diagrams can compensate the factor $a^2$
in front of them, thus giving rise to finite contributions to the $Z$'s.
Since in actual Monte Carlo simulations each $\psi$ or
$\overline{\psi}$ field is subjected to the whole rotation (\ref{eq:rotation})
\cite{err1},
the full expression (\ref{eq:o2impr}) must be considered in the calculation of
renormalization constants and mixing coefficients.

In the presentation of the computations in Sect.~8 we have for convenience
separated
all these various contributions. We will thus consider in turn diagrams with
the
insertion of the non-rotated (non-improved) operator, diagrams with the
insertion of only one rotation, either on $\psi$ or on $\overline{\psi}$,
and diagrams with the insertion of both rotations. The operator
(\ref{eq:o2impr}) has been expanded in powers of $g_0$ up to order $g_0^2$.
This
is sufficient for our 1-loop calculations. The corresponding formulae have been
computed by hand and then checked by means of the algebraic manipulation
language Schoonschip, and are summarized in appendix A.

Consequently to the use of improvement, there is a very large
number of diagrams to be computed (see appendix B) and, a part from few cases,
the algebraic manipulations of every diagram give rise to a huge number of
terms. For this reason it has been necessary to check independently all the
computations by using Schoonschip. We have to this end developed a general
program that is able to
automatically carry out all the algebraic manipulations starting from the
elementary building blocks of the calculation, represented by the expressions
of propagators and vertices. The final output of the program is an
analytic expression of the $Z$'s under the form of a 1-loop integral.

A large part of this work has thus been spent in developing an efficient
Schoon\-schip code adapted to the problem at hand. The main difficulty
resides in the fact that, although there are many built in instructions to
deal with gamma matrices, Schoonschip has been conceived having in mind
a continuum theory, which is invariant with respect to the (euclidean) Lorentz
group. On the lattice, on the contrary, the theory is only
invariant with respect to the hypercubic group, and unfortunately many simple
and common terms like
$\sum_{\lambda} \gamma_{\lambda} p_{\lambda} \sin k_{\lambda}$
that arise on the lattice are not properly handled. Actually, the above term
is wrongly reduced by Schoonschip to
$\not\!\!{p} \sin k_{\lambda}$, because two equal indices are by default
assumed to be contracted. Furthermore
the result depends on the order in which the various factors are encountered,
because the Lorentz invariant summation is immediately performed between the
first two indexed quantities encountered in the analytic expression.

Being impossible to directly use the ``gammatric'' of
Schoonschip as it is, a special routine has been developed to correctly
treat gamma matrices on the lattice while using as much as possible of
the built in Schoonschip commands\footnote{We are willing to send our codes to
anyone interested.}.

As we said, for each contribution the program starts with the expression of the
corresponding Feynman diagram in terms of propagators and vertices.
Before carrying out the Dirac algebra, one can simplify the expressions
by noticing that in the final
loop integration one will have to integrate products of sines with functions of
cosines and of H(4) invariant combinations of sines. Odd products
of sines integrate to zero, while even products will lead to integrals of the
type
\be
{\cal I}(\mu_1,\ldots,\mu_{2n}) =
\int\! d^{4} k f(\cos k, \sum_{\lambda} \sin^2 k_{\lambda})
\prod_{i=1}^{2n} \sin k_{\mu_{i}} \label{eq:sines},
\ee
where $f$ is an H(4) covariant function. Exploiting the symmetry of the
integration measure under H(4), the integrand (\ref{eq:sines}) can be
expressed in terms
of a certain number of simpler integrals. For instance for $n=2$ one has
\bd
{\cal I}(\mu_1,\mu_2,\mu_3,\mu_4) =
\int\! d^{4} k \, f(\cos k, \sum_{\lambda} \sin^2 k_{\lambda})
\sin k_{\mu_1} \sin k_{\mu_2} \sin k_{\mu_3}
\sin k_{\mu_4} =
\ed
\bd
\int\! d^{4} k \, f(\cos k, \sum_{\lambda} \sin^2 k_{\lambda})
\sin^2 k_{\mu_1} \sin^2 k_{\mu_3}
\: \delta_{\mu_1 \mu_2} \: \delta_{\mu_3 \mu_4}
\ed
\bd
+\int\! d^{4} k \, f(\cos k, \sum_{\lambda} \sin^2 k_{\lambda})
\sin^2 k_{\mu_1} \sin^2 k_{\mu_2}
\: \delta_{\mu_1 \mu_3} \: \delta_{\mu_2 \mu_4}
\ed
\bd
+\int\! d^{4} k \, f(\cos k, \sum_{\lambda} \sin^2 k_{\lambda})
\sin^2 k_{\mu_1} \sin^2 k_{\mu_2}
\: \delta_{\mu_1 \mu_4} \: \delta_{\mu_2 \mu_3}
\ed
\be
- 2 \cdot \int\! d^{4} k\, f(\cos k, \sum_{\lambda} \sin^2 k_{\lambda})
\sin^4 k_{\mu_1}
\: \delta_{\mu_1 \mu_2 \mu_3 \mu_4} \label{eq:expsines},
\ee
where $\delta_{\mu_1 \mu_2 \mu_3 \mu_4}$ is non-zero only if all the indices
are equal.

The case $n = 1$ is trivial. A 6-sine term ($n = 3$) instead gives rise to a
combination of 31 terms. In the calculation of
$<q| (O^q_{\mu \nu \tau})^{IMPR} |q>$ \cite{beccarini}
one also encounters some 8-sine
terms ($n = 4$), each one giving rise to 379 terms. A brief discussion of these
expansions and of their derivation is given in appendix C.

The rather complicated Dirac algebra (we have products of up to 7 gamma
matrices) is carried out at this point, exploiting the $\delta$-functions
that are expected to come out from the loop integration, as exemplified in
eq.~(\ref{eq:expsines}).
The last step consists in extracting the $Z$ factors by projecting the whole
expression on the appropriate tensor structure (see eqs.~(\ref{eq:ref3})).

The CPU time needed to perform the whole analytic
calculation varies according to the complexity of the diagram, going from
20 seconds up to 5 minutes for the most complicated cases on
a Sun 3 workstation.

\section{Loop integration}

The resulting analytic expressions must be finally integrated in momentum space
over the first Brillouin zone. We want to compute all these integrals with a
total error of about 1\%, in order not to spoil the accuracy aimed at with the
use of improvement.

In almost all diagrams there will
be an extra integration over a Feynman parameter $\alpha$. Since it turned out
to be exceedingly time consuming to evaluate our five dimensional integrals
with an error of less than 1\%, we decided
to perform analytically the $\alpha$ integration, writing the result
in terms of the functions
\be
{\mbox{\large F}}_{n m} \bigg( f(k),g(k)\bigg) = \int_{0}^{1} \! d\alpha \:
\frac{\alpha^{n}}{[f(k) + \alpha g(k)]^{m}} .
\ee
These functions, for the values of $n$ and $m$ relevant for this work, are
tabulated in appendix D. The remaining four dimensional integrals are
afterwards
numerically computed with the most naive rectangular method, i.e.
by summing over a uniformly distributed net of points.

All integrals are at most logarithmically divergent
in the limit $a \mu \rightarrow 0$ and have been computed
by adding and subtracting to them known integrals with integrand functions
having identical leading ``$1 / k^2$'' and ``$1 / k^4$'' lattice behavior,
as the original integrands. The final step consists in evaluating numerically
the resulting finite integrals. In this way one can easily obtain the required
precision with reasonable computing times.

As for the ``$1 / k^2$'' part, one subtracts and adds with the appropriate
coefficients the integral
\be
\int_{-\pi}^{\pi} \frac{d^4 k}{(2 \pi)^4} \cdot
\frac{1}{[4 \sum_{\lambda} \sin^2 \frac{k_{\lambda}}{2}]} = Z_0 ,
\ee
whose value can be found in Table 7.1.

\vspace{0.5 cm}
\begin{center}
\begin{tabular}{|c|c|}
\hline $Z_0$ & 0.15493339 \\
\hline $Z_1$ & 0.10778131 \\
\hline $F_0$ & 4.36922523 \\
\hline $\gamma_E$ & 0.57721566 \\
\hline
\end{tabular}
\end{center}
\begin{center}
{\small {\bf Table 7.1} - Values of some useful lattice constants}
\end{center}

For the logarithmically divergent part we actually use a double subtraction.
To explain the method, let us consider a typical integral of the form
\be
I =  \int_0^1 d\alpha \int_{-\pi}^{\pi}
\frac{d^4 k}{(2 \pi)^4} \frac{N(k)}{[D(k;\alpha)+ \alpha (1 - \alpha)
\mu^2 a^2]^2} \ \ , \ \ N(0) \neq 0 \label{eq:i} ,
\ee
where
\be
D(k;\alpha) = 4 \alpha \sum_{\lambda} \sin^2
\frac{k_{\lambda}}{2} + (1 - \alpha)[\sum_{\lambda} \sin^2 k_{\lambda} +
4 r^2 (\sum_{\lambda} \sin^2 \frac{k_{\lambda}}{2})^2]
\ee
and, for instance,
\be
N(k) = \cos k_{\mu}.
\ee
$I$ could be computed by summing and subtracting to it the known
integral
\bd
I_0 = \int_0^1 d\alpha \int_{-\pi}^{\pi} \frac{d^4 k}{(2 \pi)^4}
\frac{N(0)}{[4 \sum_{\lambda} \sin^2 \frac{k_{\lambda}}{2}
+ \alpha (1 - \alpha) \mu^2 a^2]^2}
\ed
\be
= \frac{N(0)}{16\pi^2} \cdot \Bigg[- \log \mu^2 a^2 - \int_0^1 d\alpha
\log \alpha (1 - \alpha) - \gamma_E + F_0 \Bigg] + O (a \mu) \label{eq:izero},
\ee
where $F_0$ is given in Table 7.1 together with the Eulero-Mascheroni
constant $\gamma_E$. In this way one can write
\be
I = (I - I_0) + \frac{N(0)}{16\pi^2} \cdot \Bigg[ -\log \mu^2 a^2 + 2
- \gamma_E + F_0 \Bigg] + O (a \mu) \label{eq:imenoizer}.
\ee
$I - I_0$ is finite as $a \mu \rightarrow 0$ and can be safely evaluated in
this limit. However, the
denominators in (\ref{eq:i}) and (\ref{eq:izero}) are different, and this can
lead to numerical errors larger than 1\%, if one does not use a sufficiently
large number of points, because the matching of divergent parts may not be
sufficiently good in correspondence to the smallest value of $k$ in the mesh of
points one has taken.

The results presented in literature for this kind of integrals generally use
Monte Carlo integrations. With this method one can hardly explore the
\linebreak low $k$
region with an adequate number of samplings, therefore there is a (small (?))
finite rest overlooked by the numerical integration.
To overcome this problem (which is also present, though to a lesser extent,
in the rectangle integration) and to reduce the number of different integrals
to be computed numerically, we have employed a more refined subtraction
procedure.

The idea is to perform a double subtraction in order to match separately
the form of the numerators and of the denominators appearing in
the various integrals. To do this, we introduce the auxiliary quantity
\be
I_1 = \int_0^1 d\alpha \int_{-\pi}^{\pi}
\frac{d^4 k}{(2 \pi)^4} \frac{N(0)}{[D(k;\alpha)]^2}
\ee
and we write
\be
I = I_0 + ( I_1 - I_0 ) + ( I - I_1 ) \label{eq:double1}.
\ee
Of course both differences, $I_1 - I_0$ and $I - I_1$, are finite as
$a \mu \rightarrow 0$ and can be computed directly at $a \mu = 0$.
Explicitly one gets
\be
\lim_{a\mu \rightarrow 0} (I - I_1) = \int_0^1 d\alpha \int_{-\pi}^{\pi}
\frac{d^4 k}{(2 \pi)^4} \frac{N(k) - N(0)}{[D(k;\alpha)]^2}
\label{eq:sottrfir}
\ee
and
\bd
\lim_{a\mu \rightarrow 0} (I_1 - I_0) =
N(0) \cdot \int_0^1 d\alpha \int_{-\pi}^{\pi}
\frac{d^4 k}{(2 \pi)^4} \Bigg[ \frac{1}{[D(k;\alpha)]^2}
- \frac{1}{[4 \sum_{\lambda} \sin^2 \frac{k_{\lambda}}{2}]^2} \Bigg]
\ed
\be
= N(0) \cdot \int_0^1 d\alpha \int_{-\pi}^{\pi} \frac{d^4 k}{(2 \pi)^4} \cdot
\label{eq:sottrsec}
\ee
\bd
\cdot \Bigg[ \frac{[4 \sum_{\lambda} \sin^2 \frac{k_{\lambda}}{2}]^2
- [4 \alpha \sum_{\lambda} \sin^2
\frac{k_{\lambda}}{2} + (1 - \alpha)[\sum_{\lambda} \sin^2 k_{\lambda} +
4 r^2 (\sum_{\lambda} \sin^2 \frac{k_{\lambda}}{2})^2]]^2 }
{[4 \alpha \sum_{\lambda} \sin^2
\frac{k_{\lambda}}{2} + (1 - \alpha)[\sum_{\lambda} \sin^2 k_{\lambda} +
4 r^2 (\sum_{\lambda} \sin^2 \frac{k_{\lambda}}{2})^2]]^2
[4 \sum_{\lambda} \sin^2 \frac{k_{\lambda}}{2}]^2} \Bigg].
\ed
\newline
We see that in eq.~(\ref{eq:sottrfir}) the two integrands have the same
denominator, while in eq.~(\ref{eq:sottrsec}) the two integrands have the same
numerator. The cancellation of the logarithmic divergences, between the two
terms of eq.~(\ref{eq:sottrsec}), has been made explicit in the last equality
by the exact compensation of the leading $1 / k^4$ behaviors of the two
integrands.

The double subtraction method we have explained reduces to only five the number
of different types of integrals that are in the end necessary to express all
the differences, similar to those appearing in eq.~(\ref{eq:double1}),
that arise in the diagrams one needs to compute.

In the case of diagrams which contain only gluons, we have used
the method suggested by Caracciolo, Menotti e Pelissetto \cite{pis2}
which allows to iteratively reduce all integrals to few basic integrals
which are known with great precision.

\section{Results}

\begin{figure}
\begin{center}
\begin{tabular}{|c|c|r|r|r||r|}
\hline  & {\sl r} & Wilson & $O(a)$ impr. & $O(a^2)$ impr. & total \\
\hline  & {\sl 0.2} & -1.970 & -1.746 & -0.026 & -3.742 \\
  & {\sl 0.4}& -3.081 & -4.014 & -0.173 & -7.267 \\
  {\bf SAILS} & {\sl 0.6}& -3.935 & -5.524 & -0.487 & -9.946 \\
  & {\sl 0.8} & -4.576 & -6.458 & -0.972 & -12.007 \\
  & {\sl 1.0}& -5.077 & -7.041 & -1.626 & -13.744 \\
\hline  & {\sl 0.2} & 0.555 & -0.132 & 0.017 & 0.440 \\
  & {\sl 0.4} & 1.038 & -0.470 & 0.109 & 0.678 \\
  {\bf VERTEX} & {\sl 0.6} & 1.535 & -0.821 & 0.290 & 1.004 \\
  & {\sl 0.8} & 1.952 & -1.128 & 0.555 &  1.379 \\
  & {\sl 1.0} & 2.293 & -1.389 & 0.904 & 1.808 \\
\hline  & {\sl 0.2} & -6.102 & -0.611 & 0 & -6.713 \\
  & {\sl 0.4} & -4.326 & -1.513 & 0 & -5.839 \\
  {\bf $\frac{1}{2}$ SELF-ENERGY} & {\sl 0.6} & -2.762 & -2.311 & 0 & -5.073 \\
  & {\sl 0.8} & -1.465 & -3.013 & 0 & -4.479 \\
  & {\sl 1.0} & -0.381 & -3.646 & 0 & -4.027 \\
\hline  & {\sl 0.2} & 0 & 0 & 0.006 & 0.006 \\
  {\bf operator TADPOLE} & {\sl 0.4} & 0 & 0 & 0.024 & 0.024 \\
  $+$ & {\sl 0.6} & 0 & 0 & 0.053 & 0.053 \\
  {\bf $\frac{1}{2}$ leg TADPOLE} & {\sl 0.8} & 0 & 0 & 0.095 & 0.095 \\
  & {\sl 1.0} & 0 & 0 & 0.148 & 0.148 \\
\hline \hline & {\sl 0.2} & -7.517 & -2.489 & -0.003 & -10.010 \\
  & {\sl 0.4} & -6.369 & -5.996 & -0.040 & -12.404 \\
  {\bf TOTAL} & {\sl 0.6} & -5.161 & -8.657 & -0.143 & -13.961 \\
  & {\sl 0.8} & -4.090 & -10.600 & -0.322 & -15.012 \\
  & {\sl 1.0} & -3.165 & -12.076 & -0.575 & -15.816 \\
\hline
\end{tabular}
\end{center}
\begin{center}
{\small {\bf Table 8.1} - Values of $B_{qq}$}
\end{center}
\end{figure}

\begin{figure}
\begin{center}
\begin{tabular}{|c|r|r|r||r|}
\hline   {\sl r} & Wilson & $O(a)$ impr. & $O(a^2)$ impr. & total \\
\hline   {\sl 0.2} & -5.184 & -0.029 & -0.010 & -5.223 \\
   {\sl 0.4} & -1.609 & -0.298 & -0.038 & -1.945 \\
   {\sl 0.6} & -0.704 & -0.582 & -0.063 & -1.349 \\
   {\sl 0.8} & -0.276 & -0.782 & -0.100 & -1.158 \\
   {\sl 1.0} & 0.019 & -0.906 & -0.154 & -1.041 \\
\hline
\end{tabular}
\end{center}
\begin{center}
{\small {\bf Table 8.2} - Values of $B_{qg}$}
\end{center}
\begin{center}
\begin{tabular}{|c|r|r||r|}
\hline   {\sl r} & Wilson & $O(a)$ impr. & total \\
\hline   {\sl 0.2} & -3.722 & 0.217 & -3.505 \\
   {\sl 0.4} & -4.372 & 0.628 & -3.744 \\
   {\sl 0.6} & -4.938 & 1.044 & -3.894 \\
   {\sl 0.8} & -5.413 & 1.427 & -3.986 \\
   {\sl 1.0} & -5.817 & 1.773 & -4.044 \\
\hline
\end{tabular}
\end{center}
\begin{center}
{\small {\bf Table 8.3} - Values of $B_{gq}$}
\end{center}
\begin{center}
\begin{tabular}{|c|l|}
\hline  & Wilson \\
\hline  {\bf SAILS} & -4.453 \\
\hline  {\bf VERTEX} & \ 5.019 \\
\hline  {\bf operator TADPOLE} & -40.024 +4$\pi^{2}/N_c^2$ \\
\hline  {\bf $\frac{1}{2}$ leg TADPOLE} & \ 19.207 -2$\pi^{2}/N_c^2$ \\
\hline  {\bf $\frac{1}{2}$ gluon LOOP} & \ 2.274  \\
\hline  {\bf $\frac{1}{2}$ GHOST} & \ 0.198  \\
\hline \hline {\bf TOTAL} & -17.778 +2$\pi^{2}/N_c^2$  \\
\hline
\end{tabular}
\end{center}
\begin{center}
{\small {\bf Table 8.4} - Values of $B_{gg}$}
\end{center}
\end{figure}
\begin{figure}
\begin{center}
\begin{tabular}{|c|c|}
\hline
\hline  & \\ {\bf SAILS} & $ {\displaystyle\frac{1}{192}}
- {\displaystyle\frac{7}{9\pi^2}}
+ {\displaystyle\frac{31}{24}Z_0}
- {\displaystyle\frac{19}{48}Z_1}
- {\displaystyle\frac{7}{24\pi^2} \cdot \bigg( F_0-\gamma_E \bigg)}$  \\ & \\
\hline  & \\ {\bf VERTEX} & $ - {\displaystyle\frac{13}{192}}
+ {\displaystyle\frac{23}{48\pi^2}}
- {\displaystyle\frac{53}{144}Z_0}
+ {\displaystyle\frac{1}{3}Z_1}
+ {\displaystyle\frac{3}{16\pi^2} \cdot \bigg( F_0-\gamma_E \bigg)}$  \\ & \\
\hline  & \\ {\bf operator TADPOLE} & $ - {\displaystyle\frac{3}{64}}
- {\displaystyle\frac{4}{3}Z_0}
+ {\displaystyle\frac{1}{4 N_c^2}}$  \\ & \\
\hline  & \\ {\bf $\frac{1}{2}$ leg TADPOLE} & $ {\displaystyle\frac{1}{32}}
+ {\displaystyle\frac{7}{12}Z_0}
- {\displaystyle\frac{1}{8 N_c^2}}$  \\ & \\
\hline  & \\ {\bf $\frac{1}{2}$ gluon LOOP} & $ {\displaystyle\frac{1}{32}}
+ {\displaystyle\frac{13}{72\pi^2}}
- {\displaystyle\frac{17}{36}Z_0}
+ {\displaystyle\frac{19}{192\pi^2} \cdot \bigg( F_0-\gamma_E \bigg)}$  \\ & \\
\hline  & \\ {\bf $\frac{1}{2}$ GHOST} & $ {\displaystyle\frac{1}{72\pi^2}}
- {\displaystyle\frac{1}{72}Z_0}
+ {\displaystyle\frac{1}{192\pi^2} \cdot \bigg( F_0-\gamma_E \bigg)}$  \\ & \\
\hline \hline & \\ {\bf TOTAL} & $ -{\displaystyle\frac{3}{64}}
- {\displaystyle\frac{5}{48\pi^2}}
- {\displaystyle\frac{5}{16}Z_0}
- {\displaystyle\frac{1}{16}Z_1}
+ {\displaystyle\frac{1}{8 N_c^2}}$ \\ & \\
\hline \hline
\end{tabular}
\end{center}
\begin{center}
{\small {\bf Table 8.5} - Analytic expressions of $B_{gg} / (16 \pi^2)$. The
constants $Z_0$, $Z_1$, $F_0$ and $\gamma_E$ can be found in Table 7.1.
Each line in this Table equals the sum of the corresponding two in Tables
8.7 and 8.8.}
\end{center}
\end{figure}
\begin{figure}
\begin{center}
\begin{tabular}{|c|c|r|r||r|}
\hline  & {\sl r} & Wilson & $O(a)$ impr. & total \\
\hline & {\sl 0.2} & -13.473 & -27.475  & -40.948  \\
  & {\sl 0.4} & -5.643  & -32.141
  & -37.784  \\
  {\bf $\frac{1}{2}$ quark LOOP} & {\sl 0.6} & -3.512 & -27.057  &
  -30.570  \\
  & {\sl 0.8} & -2.656  & -21.581  & -24.237  \\
  & {\sl 1.0} & -2.168  & -17.257  & -19.425  \\
\hline
\end{tabular}
\end{center}
\begin{center}
{\small {\bf Table 8.6} - Values of $B^{f}_{gg}$}
\end{center}
\end{figure}

The results of our calculations are summarized in the Tables 8.1 to 8.6
presented in this section. The contributions coming from the standard Wilson
action (that is, the non-improved results), those coming from the terms of
order $a$ in the improvement
and those coming from the terms of order $a^2$ are separately
shown. Contributions coming from different classes of diagrams, according
to the classification given in appendix B, are also separately presented. In
Tables 8.7 and 8.8 we detail the comparison of our results with those
of ref.~\cite{pis2}.

In the quenched approximation ($N_f = 0$) $B_{qg}$ and $B^f_{gg}$ disappear
from
eqs.~(\ref{eq:eigve1}) and (\ref{eq:eigve2}), and even in the flavor singlet
sector the quark operator does not mix with the gluon operator, as the matrix
(\ref{eq:matrix}) becomes triangular. In the full theory
($N_f \ne 0$, ``unquenched'') we have on the contrary complete mixing
between the quark operator, $O^q$, and the gluon operator, $O^g$.
To compute in this case the renormalization constants and mixing coefficients
we need also to evaluate the diagrams of Figs.~6 and 7 of appendix B,
corresponding to respectively the matrix element
$<g,\sigma| (O^q_{\mu \nu})^{IMPR} |g,\sigma>$ and the quark loop
contribution to the matrix element $<g,\sigma| O^g_{\mu \nu} |g,\sigma>$.

In Table 8.5 one can find the analytic expression of $B_{gg}$ leading,
through the use of Table 7.1, to the numerical values shown in Table 8.4.

The contributions to $<g,\sigma| O^g_{\mu \nu} |g,\sigma>$
of the diagrams labeled as ``tadpo\-le-QCD vertex'' and
``tadpole-quark loop'' in appendix B
are zero, and have not been inserted in Tables 8.4 and 8.5.

As emphasized in the previous section, the accuracy of our results is
better than 1\%. The differences we have found by comparing our
results for the Wilson case with results previously appeared in the literature
\cite{kronfeld,Mart1,Mart2,Ross} come from either
insufficient accuracy in the Monte Carlo numerical integration or trivial
algebraic mistakes.
However, we are in complete agreement with the results for the energy-momentum
tensor given in refs.~\cite{pis1} and \cite{pis2}, as well
as with the gluon self-energy reported in \cite{kawai}.

To check the value of $Z_{qq}$, it is enough to take the corresponding
quantity, called $Z_1$ in Table 1 of ref.~\cite{pis1}, and multiply it by
$-8$ and $16 \pi^2$ to find
\be
B_{qq}~\mbox{(from ref.~\cite{pis1})} = -8 \cdot 16 \pi^2 \cdot ( 0.00252 ) ,
\ee
in agreement (within numerical accuracy) with the number given in the last
line of Table~8.1:
\be
B_{qq} = -3.165.
\ee

To compare the results for $Z_{gg}$ with those of ref.~\cite{pis2} one must
take into account that two different regularization procedures have been
employed to deal with logarithmically divergent loop integrals. We have
regularized all integrals by putting the external legs infinitesimally
off-shell by an amount $ \mu^2 \ll 1 / a^2 $, while in ref.~\cite{pis2} each
term, $I$, is expanded up to second order in the four-vector of the external
legs, $p$, and decomposed in H(4) tensors. Dimensional regularization is used
to compute loop integrals.
Calling $J$, as in ref.~\cite{pis2}, the relevant terms of this expansion,
it is easy to see that the difference $I-J$ (for small $p$) can in fact be
computed as a continuum integral\footnote{This is why only the genuinely
lattice dependent quantity $J$ is reported in ref.~\cite{pis2}.},
since its value only depends on the behavior of the integrand at $k=0$.

To explicitly carry on the comparison with ref.~\cite{pis2} we have
thus computed using dimensional regularization $I-J$ for small $p$.
We report their values in Table 8.7, apart from the logarithmic terms we always
have singled out in all formulae. We collect in Table 8.8 the values of $J$,
obtained from ref.~\cite{pis2} remembering that in the Tables presented there
our operator (\ref{eq:og}) corresponds in their notations to the tensor
structure $L_3 - L_{10}$. As announced, the sum $J + (I-J)$ reproduce exactly
the results given in our Table 8.5.

\begin{figure}
\begin{center}
\begin{tabular}{|c|c|}
\hline
\hline  & \\ {\bf SAILS} &$ - {\displaystyle\frac{7}{24\pi^2}
\cdot \bigg( \frac{2}{\varepsilon} +\log 4\pi -\gamma_E \bigg)}$  \\ & \\
\hline  & \\ {\bf VERTEX} &$ {\displaystyle\frac{7}{24\pi^2}}
+ {\displaystyle\frac{3}{16\pi^2}
\cdot \bigg( \frac{2}{\varepsilon} +\log 4\pi -\gamma_E \bigg)}$  \\ & \\
\hline  & \\ {\bf operator TADPOLE} & $0$ \\ & \\
\hline  & \\ {\bf $\frac{1}{2}$ leg TADPOLE} & $0$ \\ & \\
\hline  & \\ {\bf $\frac{1}{2}$ gluon LOOP} &$ {\displaystyle\frac{29}
{144\pi^2}} + {\displaystyle\frac{19}{192\pi^2}
\cdot \bigg( \frac{2}{\varepsilon} +\log 4\pi -\gamma_E \bigg)}$ \\ & \\
\hline  & \\ {\bf $\frac{1}{2}$ GHOST} &$ {\displaystyle\frac{1}{72\pi^2}}
+ {\displaystyle\frac{1}{192\pi^2}
\cdot \bigg( \frac{2}{\varepsilon} +\log 4\pi -\gamma_E \bigg)}$  \\ & \\
\hline \hline & \\ {\bf TOTAL $I - J$}
&$ {\displaystyle\frac{73}{144\pi^2}}  $\\ & \\
\hline \hline
\end{tabular}
\end{center}
{\small {\bf Table 8.7} - The contribution to the analytic expression of
$B_{gg} / (16 \pi^2)$, called $I - J$ in the text, according to the definition
given in ref.~\cite{pis2}.}
\end{figure}

\begin{figure}
\begin{center}
\begin{tabular}{|c|c|}
\hline
\hline  & \\ {\bf SAILS} & $ {\displaystyle\frac{1}{192}}
- {\displaystyle\frac{7}{9\pi^2}}
+ {\displaystyle\frac{31}{24}Z_0}
- {\displaystyle\frac{19}{48}Z_1}
- {\displaystyle\frac{7}{24\pi^2}
\cdot \bigg( - \frac{2}{\varepsilon} + F_0 -\log 4\pi \bigg)}$ \\ & \\
\hline  & \\ {\bf VERTEX} & $ - {\displaystyle\frac{13}{192}}
+ {\displaystyle\frac{3}{16\pi^2}}
- {\displaystyle\frac{53}{144}Z_0}
+ {\displaystyle\frac{1}{3}Z_1}
+ {\displaystyle\frac{3}{16\pi^2}
\cdot \bigg( - \frac{2}{\varepsilon} + F_0 -\log 4\pi \bigg)}$ \\ & \\
\hline  & \\ {\bf operator TADPOLE} & $ - {\displaystyle\frac{3}{64}}
- {\displaystyle\frac{4}{3}Z_0}
+ {\displaystyle\frac{1}{4 N_c^2}}$ \\ & \\
\hline  & \\ {\bf $\frac{1}{2}$ leg TADPOLE} & $
{\displaystyle\frac{1}{32}} + {\displaystyle\frac{7}{12}Z_0}
- {\displaystyle\frac{1}{8 N_c^2}}$ \\ & \\
\hline  & \\ {\bf $\frac{1}{2}$ gluon LOOP} &
$ {\displaystyle\frac{1}{32}}
- {\displaystyle\frac{1}{48\pi^2}}
- {\displaystyle\frac{17}{36}Z_0}
+ {\displaystyle\frac{19}{192\pi^2}
\cdot \bigg( - \frac{2}{\varepsilon} + F_0 -\log 4\pi \bigg)}$ \\ & \\
\hline  & \\ {\bf $\frac{1}{2}$ GHOST} & $
- {\displaystyle\frac{1}{72}Z_0}
+ {\displaystyle\frac{1}{192\pi^2}
\cdot \bigg( - \frac{2}{\varepsilon} + F_0 -\log 4\pi \bigg)}$ \\ & \\
\hline \hline & \\ {\bf TOTAL $J$} & $ -{\displaystyle\frac{3}{64}}
- {\displaystyle\frac{11}{18\pi^2}}
- {\displaystyle\frac{5}{16}Z_0}
- {\displaystyle\frac{1}{16}Z_1}
+ {\displaystyle\frac{1}{8 N_c^2}}$ \\ & \\
\hline \hline
\end{tabular}
\end{center}
{\small {\bf Table 8.8} - The contribution to the analytic expression of
$B_{gg} / (16 \pi^2)$, called $J$ in the text, according to the definition
given in ref.~\cite{pis2}.}
\end{figure}

Obviously physical quantities will be at the end independent from the
chosen regularization procedure. In particular the dependence from
the subtraction point, $\mu$, must disappear from physical hadronic matrix
elements. In fact, consistently to each order in perturbation theory, the
$\log \mu$ terms get canceled in the product between the Wilson coefficients
and the matrix elements of the
renormalized operators that are eigenvectors of the anomalous dimension matrix.
The net result is that effectively the Wilson coefficients must be taken at a
momentum scale $a^{-1}$ and the operators (\ref{eq:eigve1}) and
(\ref{eq:eigve2}) renormalized by reduced renormalization constants, obtained
   from the full expressions by dropping all logarithmic terms.

The effective renormalization of the bare quark operator is rather small.
Numerically for $N_c = 3$ at the typical values
$\beta \equiv 2 N_c / g_0^2 = 6$ and $r = 1$, one gets in the quenched
approximation for the unmixed quark operator
\begin{center}
\begin{tabular}{rll}
$\widetilde{O}^2_{\mu \nu} =$&$ 1.027~O^q_{\mu \nu} $& Wilson case \\
$(\widetilde{O}^2_{\mu \nu})^{IMPR} =$&$ 1.134~(O^q_{\mu \nu})^{IMPR} $&
Improved case ,
\end{tabular}
\end{center}
where by the superscript $\widetilde{}$ we mean that the $\log \mu$ term
contribution has been dropped from the expression of the $Z$'s with the
understanding that the corresponding Wilson
coefficients are computed at a scale $\mu = 1 / a$.

The gluonic contribution to the energy-momentum tensor operator, whose
hadro\-nic matrix elements are incidentally very difficult to measure in
Monte Carlo simulations \cite{Mart1}, are somewhat larger. One gets in fact
for $\widehat{O}^1$
\begin{center}
\begin{tabular}{rll}
$\widehat{O}^1_{\mu \nu} =$&$ 1.296~O^g_{\mu \nu} - 0.978~O^q_{\mu \nu} $&
Wilson case \\
$(\widehat{O}^1_{\mu \nu})^{IMPR} =$&$ 1.296~O^g_{\mu \nu}
- 1.099~(O^q_{\mu \nu})^{IMPR} $& Improved case .
\end{tabular}
\end{center}

There are suggestions that the large renormalizations in the gluon sector
come from the choice of the
bare lattice coupling as expansion parameter \cite{parisi}. Actually, a very
large renormalization factor relates the bare lattice coupling to the
continuum one (as calculated for example in the $\overline{MS}$ scheme
\cite{hasenfratz}).
It is also believed that the large contributions coming from tadpole diagrams
are the main source of numerical discrepancies between continuum and lattice
perturbation theory.

According to ref.~\cite{tadimp}, a resummation of tadpole contributions could
be
effectively achieved if the perturbative series is expanded in terms of a
``boosted'' coupling constant
\be
\widetilde{g} = \frac{g_0}{u_0^2} \label{eq:boosted},
\ee
where $u_0 \: (< 1)$ should be taken as the non-perturbative mean value of the
link operator. A convenient measure of $u_0$ can be obtained from the plaquette
expectation value (eq.~\ref{eq:effeno})
\be
u_0 = \left< \frac{1}{3}~\mbox{Tr} U_{n,\mu \nu} \right>^{1/4}
\label{eq:quapl}.
\ee
Here a first question arises whether or not tadpole diagrams should be
retraced from perturbative results, as they are effectively already taken into
account in the redefinition (\ref{eq:boosted}).

Another problem is that $u_0$ could also be measured from the renormalization
of the Wilson hopping parameter, by writing
\be
K_c= \frac{1}{8 \cdot u_0} \label{eq:kcritico}.
\ee
The definition (\ref{eq:kcritico}), contrary to the one given by
eq.~(\ref{eq:quapl}), depends upon the choice of the fermion part of the QCD
action\footnote{It is interesting to note that, unlike the Wilson case, the
two definitions (\ref{eq:quapl}) and (\ref{eq:kcritico}) seem to lead to
numerically very close values in the case of the ``clover-leaf'' action
\cite{priv}.}.
It should be observed here that, when using eq.~(\ref{eq:kcritico}), one should
for consistency perhaps also introduce an explicit correction for the large
deviation between the perturbative and the non-perturbative quark mass
renormalization\footnote{We wish to thank G.Martinelli for a discussion
about this point.}, besides changing the value of the effective expansion
parameter.

Because of the many subtleties involved in the whole question of ``boosting''
perturbation theory, we like to refrain from attempting any numerical
evaluation of the effects of these corrections to our numbers, and we rather
like to present the results of our calculations as they appears in terms of
the bare coupling constant, leaving any other numerical consideration
to actual applications.

\section{Conclusions and outlook}

As we have briefly discussed in Sect.~3, a simple modification of the fermion
action, accompanied by a suitable redefinition of hadronic operators, leads to
the theoretical expectation of an ``effective $O(a)$'' improvement in on-shell
hadronic matrix elements \cite{rom1}. This was indeed beautifully confirmed by
the exploratory numerical Monte Carlo simulations of refs.~\cite{err1} and
\cite{err2}.

After these encouraging results there has been a general agreement that the
new high-statistics QCD Monte Carlo simulations that are planned to use
next generation dedicated machines (e.g. APE with $\geq 6$ Gflops) should
all be performed with the ``clover-leaf'' improved action.

This paper and the forthcoming one, that are dealing with respectively the
first and the second moment of DIS structure functions, are part of the
general project aimed at the perturbative recalculation with the
``clover-leaf''
improved action of all the renormalization constants and mixing coefficients
of the fermion operators whose matrix elements are phenomenologically relevant
and susceptible of being measured in Monte Carlo simulations.

To complete the analytic part of the above program only the mixing coefficient
between the dimension six four-fermion $\Delta I = 1 / 2$ effective weak
Hamiltonian and the dimension five operator appearing in eq.~(\ref{eq:impr})
is missing. In the standard Wilson theory this finite (thanks to
the GIM mechanism) coefficient was computed in ref.~\cite{curci}.
Its calculation in the case of the ``clover-leaf'' improved action is presently
under way \cite{capitani}.

We think it is fair to conclude this paper with a remark on the reliability
of lattice perturbation theory itself, which looks even more problematic than
continuum perturbation theory. For instance, gluonic corrections tend to be
anomalously big because of large tadpole contributions. As we remarked in
Sect.~8, one possible remedy could be to use a better expansion parameter
than the bare coupling constant, as advocated in refs.~\cite{parisi} and
\cite{tadimp}, in a way to effectively incorporate a resummation of these
large tadpole diagrams.

A perhaps more promising and consistent way to proceed has been, however,
recently suggested in ref.~\cite{nonpert}. The idea is to extract
renormalization constants directly from Monte Carlo data, by measuring matrix
elements of hadronic operators between quark states and exploiting
information coming from Ward identities. Such a procedure would eliminate
altogether lattice perturbation theory from the game. This fact might be
regarded as rather satisfactory for a numerical approach like the lattice one
which, taken in its most extreme formulation, should refrain from using any
kind of (approximate or unreliable) perturbative calculations.

\section*{Acknowledgments}

We would like to thank Sergio Caracciolo, Pietro Menotti and Andrea Pelissetto
for their patience in helping us to compare their calculations with ours.
We also thank Giuseppe Beccarini for checking some of our Schoonschip outputs.
S.C. would like to thank Guido Martinelli and Massimo Testa for introducing
him to lattice QCD, Enrico Franco, Roberto Frezzotti, Emidio Gabrielli
and Carlotta Pittori for help at the initial stage of this work and Ugo
Aglietti for useful hints about numerical integration.

\newpage

\appendix

\section*{Appendix A}

\section*{Perturbative expansion of the vertex operators
$(O^q_{\mu \nu})^{IMPR}$ and $O^g_{\mu \nu}$}

In this appendix we give the perturbative expansion of the vertex operators
$(O^q_{\mu \nu})^{IMPR}$, given by eq.~(\ref{eq:o2impr}), and $O^g_{\mu \nu}$,
given by eq.~(\ref{eq:og}) with the insertion of the definition
(\ref{eq:effesi}). We used the covariant derivatives defined as
\be
\stackrel{\displaystyle \rightarrow}{D}_{\mu} \psi_n = \frac{1}{2a}
\Big[ U_{n, \nu} \psi_{n + \mu} - U_{n - \mu, \mu}^{+} \psi_{n - \mu} \Big]
\ee
\bd
\overline{\psi} \stackrel{\displaystyle \leftarrow}{D}_{\mu}  = \frac{1}{2a}
\Big[ \overline{\psi}_{n + \mu} U_{n, \nu}^{+}
 - \overline{\psi}_{n - \mu} U_{n - \mu, \mu} \Big] ,
\ed
and the conventions
\be
A_{n, \mu} = \int\! \frac{d^4 q}{(2 \pi)^4} \:
e^{\displaystyle i (q + q_{\mu} /2) n} A_{\mu}(q)
\ee
\be
\psi_n = \int\! \frac{d^4 q}{(2 \pi)^4} \: e^{\displaystyle i q n} \psi (q)
\ee
\be
\overline{\psi}_n = \int\! \frac{d^4 q}{(2 \pi)^4} \:
e^{\displaystyle - i q n} \overline{\psi} (q) ,
\ee
where the integrals are performed over the first Brillouin zone.
Throughout this appendix external and loop momenta are expressed in lattice
units.

Since the full Fourier transform of the operator $(O^q_{\mu \nu})^{IMPR}$ is
very complicated, we give here only the form it effectively takes when inserted
in the diagrams of Figs.~2, 3, 4 and 7. Calling $p$ the external incoming and
outgoing momentum, and
$k$ the fermion loop momentum, one finds, separating the various contributions
according to their naive order in $a$:

a) tree level

\bd
(O^q_{\mu \nu})^{IMPR}(n = 0) |_{tree} =
{\displaystyle\frac{i}{2}} \int_{-\pi}^{\pi} \frac{d^4 k}{(2 \pi)^4}
\overline{\psi}(k) \gamma_{\mu} {\displaystyle\frac{\sin k_{\nu}}{a}}
\psi (k)
\ed
\be
+ {\displaystyle\frac{a r }{2}}
\int_{-\pi}^{\pi} \frac{d^4 k}{(2 \pi)^4} \overline{\psi}(k)
{\displaystyle\frac{\sin k_{\mu}}{a}} {\displaystyle\frac{\sin k_{\nu}}{a}}
\psi (k) \label{eq:qtree}
\ee
\bd
- {\displaystyle\frac{i a^2 r^2}{8}} \int_{-\pi}^{\pi} \frac{d^4 k}{(2 \pi)^4}
\overline{\psi}(k) \sum_{\lambda,\lambda '}
\gamma_{\lambda} \gamma_{\mu} \gamma_{\lambda '}
{\displaystyle\frac{\sin k_{\lambda}}{a}}
{\displaystyle\frac{\sin k_{\lambda '}}{a}}
{\displaystyle\frac{\sin k_{\nu}}{a}} \psi (k).
\ed

b) order $g_0$

\bd
(O^q_{\mu \nu})^{IMPR}(n = 0) |_{g_0} =
{\displaystyle\frac{i g_0}{2}} \int_{-\pi}^{\pi} \frac{d^4 k}{(2 \pi)^4}
\overline{\psi}(p) \gamma_{\mu} \Big[
\cos ({\displaystyle\frac{k + p}{2}})_{\nu} A_{\nu}(p-k) \Big] \psi (k),
\ed
\bd
+ {\displaystyle\frac{a g_0 r}{4}} \int_{-\pi}^{\pi} \frac{d^4 k}{(2 \pi)^4}
\overline{\psi}(p) \sum_{\lambda}
\Bigg[ \cos ({\displaystyle\frac{k + p}{2}})_{\nu}
\left( \gamma_{\mu} \gamma_{\lambda} {\displaystyle\frac{\sin k_{\lambda}}{a}}
+ \gamma_{\lambda} \gamma_{\mu} {\displaystyle\frac{\sin p_{\lambda}}{a}}
\right) A_{\nu}(p-k)
\ed
\bd
+ \cos ({\displaystyle\frac{k + p}{2}})_{\lambda}
\left( \gamma_{\mu} \gamma_{\lambda} {\displaystyle\frac{\sin p_{\nu}}{a}} +
\gamma_{\lambda} \gamma_{\mu} {\displaystyle\frac{\sin k_{\nu}}{a}}
\right) A_{\lambda}(p-k) \Bigg] \psi (k)
\ed
\bd
- {\displaystyle\frac{i a^2 g_0 r^2}{8}} \int_{-\pi}^{\pi}
\frac{d^4 k}{(2 \pi)^4} \overline{\psi}(p) \sum_{\lambda,\lambda '}
\gamma_{\lambda} \gamma_{\mu} \gamma_{\lambda '}
\Bigg[ \cos ({\displaystyle\frac{k + p}{2}})_{\nu} A_{\nu}(p-k)
{\displaystyle\frac{\sin k_{\lambda '}}{a}}
{\displaystyle\frac{\sin p_{\lambda}}{a}}
\ed
\bd
+ \cos ({\displaystyle\frac{k + p}{2}})_{\lambda} A_{\lambda}(p-k)
{\displaystyle\frac{\sin k_{\lambda '}}{a}}
{\displaystyle\frac{\sin k_{\nu}}{a}}
\ed
\be
+ \cos ({\displaystyle\frac{k + p}{2}})_{\lambda '} A_{\lambda '}(p-k)
{\displaystyle\frac{\sin p_{\lambda}}{a}}
{\displaystyle\frac{\sin p_{\nu}}{a}}
\Bigg] \psi (k)  \label{eq:qg}.
\ee
This formula is given in the kinematical configuration in which the incoming
gluon momentum lands on the incoming quark leg. If the gluon is attached to the
outgoing quark leg, one must exchange $p$ and $k$. Notice that the first lines
in eqs.~(\ref{eq:qtree}) and (\ref{eq:qg}) correspond to the non-improved
expression of the operator.

We do not give here the expression of the $O(g_0^2)$ terms because, besides
being extremely complicated, they are not actually necessary for our
computation. In fact we need them only when either the gluon or the quark legs
are contracted to make a tadpole loop, and in this situation the tadpole
directly factorizes out.

On the gluon operator $O^g_{\mu \nu}$ there is no
effect due to the improvement, because the transformation (\ref{eq:rotation})
acts only on spinors, and one finds:

a) tree level

\bd
O^g_{\mu \nu}(n = 0) |_{tree} = - {\displaystyle\frac{1}{2}}
\sum_{\rho} \int {\displaystyle\frac{dq_1 \, dq_2}{(2 \pi)^4}}
\delta^{(4)} (q_1 + q_2)
\ed
\bd
\frac{1}{a} \left( A_{\mu}^a (q_1)
\cos {\displaystyle\frac{q_{1\mu}}{2}} \sin q_{1\rho}
- A_{\rho}^a (q_1) \cos {\displaystyle\frac{q_{1\rho}}{2}} \sin q_{1\mu}
\right)
\ed
\be
\cdot \frac{1}{a} \left( A_{\rho}^a (q_2)
\cos {\displaystyle\frac{q_{2\rho}}{2}} \sin q_{2\nu}
- A_{\nu}^a (q_2) \cos {\displaystyle\frac{q_{2\nu}}{2}} \sin q_{2\rho}
\right) \label{eq:gtree} .
\ee

b) order $g_0$

\bd
O^g_{\mu \nu}(n = 0) |_{g_0} = - {\displaystyle\frac{i g_0}{4}} f^{abc}
\sum_{\rho} \int {\displaystyle\frac{dq_1 \, dq_2 \, dq_3}{(2 \pi)^8}}
\delta^{(4)} (q_1 + q_2 + q_3)
\ed
\bd
{\displaystyle\frac{1}{a}} \cdot \Bigg[ A_{\mu}^a (q_1)
\cos {\displaystyle\frac{q_{1\mu}}{2}} \sin q_{1\rho}
- A_{\rho}^a (q_1) \cos {\displaystyle\frac{q_{1\rho}}{2}} \sin q_{1\mu}
\Bigg]
\ed
\bd
\cdot \Bigg[ A_{\rho}^b (q_2) A_{\nu}^c (q_3)
\Big[ \left( \cos {\displaystyle\frac{q_{2\rho}}{2}}
- \cos {\displaystyle\frac{(q_2 + 2 q_3)_{\rho}}{2}} \right)
\cdot \left( \cos {\displaystyle\frac{q_{3\nu}}{2}}
- \cos {\displaystyle\frac{(q_3 + 2 q_2)_{\nu}}{2}} \right)
\ed
\bd
- 2 \cos {\displaystyle\frac{(q_2 + 2 q_3)_{\rho}}{2}}
\cos {\displaystyle\frac{(q_3 + 2 q_2)_{\nu}}{2}} \Big]
- A_{\rho}^b (q_2) A_{\rho}^c (q_3)
\sin {\displaystyle\frac{(q_2 + q_3)_{\rho}}{2}} \sin q_{3\nu}
\label{eq:gg}
\ed
\be
+ A_{\nu}^b (q_2) A_{\nu}^c (q_3)
\sin {\displaystyle\frac{(q_2 + q_3)_{\nu}}{2}} \sin q_{3\rho}
\Bigg] + (\mu \rightarrow \nu) .
\ee

As for the $O(g_0^2)$ terms we do not report here their expression for the
same reasons explained after eq.~(\ref{eq:qg}).

\section*{Appendix B}

\section*{Diagrams}

In what follows we show all the 1-loop diagrams that have been calculated
in the context of this work. Many of them have a label that allows an
easy connection with the Tables presented in Sect.~8.

Actually, each graph in this appendix corresponds to several diagrams
in the perturbative expansion. For example, the vertex correction of Fig.~2
\bd
\begin{picture}(40000,15000)
\drawline\fermion[\NE\REG](10000,2000)[3250]
\drawline\gluon[\E\REG](\pbackx,\pbacky)[9]
\drawline\fermion[\NE\REG](\fermionbackx,\fermionbacky)[6750]
\drawline\fermion[\NW\REG](\gluonbackx,\gluonbacky)[6750]
\put(\fermionbackx,\fermionbacky){\circle*{800}}
\drawline\fermion[\SE\REG](\gluonbackx,\gluonbacky)[3250]
\end{picture}
\ed
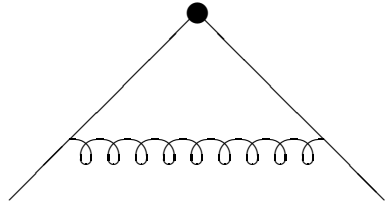
{\small {\bf Fig. 2} - The graph that symbolically represents the 1-loop
correction to the insertion of the $(O^q_{\mu \nu})^{IMPR}$ operator.
The insertion is indicated by a dot. The wavy line is a gluon.}
\vspace{1 cm}

\noindent
symbolically represents the sum of 12 different diagrams.
They are all shown in Figs.~3.1 to 3.12, where the improved quark-gluon vertex
(\ref{eq:newimpr})
is represented by a dot and the insertion of $O(a$) and $O(a^2)$ corrections
to the vertex operators (see also eqs.~(\ref{eq:o2impr}) and (\ref{eq:qtree}))
are respectively indicated by an open circle and a square.
\bd
\begin{picture}(40000,15000)
\put(0,0){3.1}
\drawline\fermion[\NE\REG](0,2000)[3250]
\drawline\gluon[\E\REG](\pbackx,\pbacky)[9]
\drawline\fermion[\NE\REG](\fermionbackx,\fermionbacky)[6750]
\drawline\fermion[\NW\REG](\gluonbackx,\gluonbacky)[6750]
\drawline\fermion[\SE\REG](\gluonbackx,\gluonbacky)[3250]
\put(20000,0){3.2}
\drawline\fermion[\NE\REG](20000,2000)[3250]
\drawline\gluon[\E\REG](\pbackx,\pbacky)[9]
\put(\fermionbackx,\fermionbacky){\circle*{500}}
\drawline\fermion[\NE\REG](\fermionbackx,\fermionbacky)[6750]
\drawline\fermion[\NW\REG](\gluonbackx,\gluonbacky)[6750]
\drawline\fermion[\SE\REG](\gluonbackx,\gluonbacky)[3250]
\end{picture}
\ed
\bd
\begin{picture}(40000,15000)
\put(0,0){3.3}
\drawline\fermion[\NE\REG](0,2000)[3250]
\drawline\gluon[\E\REG](\pbackx,\pbacky)[9]
\drawline\fermion[\NE\REG](\fermionbackx,\fermionbacky)[6750]
\drawline\fermion[\NW\REG](\gluonbackx,\gluonbacky)[6750]
\put(\gluonbackx,\gluonbacky){\circle*{500}}
\drawline\fermion[\SE\REG](\gluonbackx,\gluonbacky)[3250]
\put(20000,0){3.4}
\drawline\fermion[\NE\REG](20000,2000)[3250]
\drawline\gluon[\E\REG](\pbackx,\pbacky)[9]
\put(\fermionbackx,\fermionbacky){\circle*{500}}
\drawline\fermion[\NE\REG](\fermionbackx,\fermionbacky)[6750]
\drawline\fermion[\NW\REG](\gluonbackx,\gluonbacky)[6750]
\put(\gluonbackx,\gluonbacky){\circle*{500}}
\drawline\fermion[\SE\REG](\gluonbackx,\gluonbacky)[3250]
\end{picture}
\ed
\bd
\begin{picture}(40000,15000)
\put(0,0){3.5}
\drawline\fermion[\NE\REG](0,2000)[3250]
\drawline\gluon[\E\REG](\pbackx,\pbacky)[9]
\drawline\fermion[\NE\REG](\fermionbackx,\fermionbacky)[6750]
\drawline\fermion[\NW\REG](\gluonbackx,\gluonbacky)[6750]
\put(\fermionbackx,\fermionbacky){\circle{500}}
\drawline\fermion[\SE\REG](\gluonbackx,\gluonbacky)[3250]
\put(20000,0){3.6}
\drawline\fermion[\NE\REG](20000,2000)[3250]
\drawline\gluon[\E\REG](\pbackx,\pbacky)[9]
\put(\fermionbackx,\fermionbacky){\circle*{500}}
\drawline\fermion[\NE\REG](\fermionbackx,\fermionbacky)[6750]
\drawline\fermion[\NW\REG](\gluonbackx,\gluonbacky)[6750]
\put(\fermionbackx,\fermionbacky){\circle{500}}
\drawline\fermion[\SE\REG](\gluonbackx,\gluonbacky)[3250]
\end{picture}
\ed
\bd
\begin{picture}(40000,15000)
\put(0,0){3.7}
\drawline\fermion[\NE\REG](0,2000)[3250]
\drawline\gluon[\E\REG](\pbackx,\pbacky)[9]
\drawline\fermion[\NE\REG](\fermionbackx,\fermionbacky)[6750]
\drawline\fermion[\NW\REG](\gluonbackx,\gluonbacky)[6750]
\put(\fermionbackx,\fermionbacky){\circle{500}}
\put(\gluonbackx,\gluonbacky){\circle*{500}}
\drawline\fermion[\SE\REG](\gluonbackx,\gluonbacky)[3250]
\put(20000,0){3.8}
\drawline\fermion[\NE\REG](20000,2000)[3250]
\drawline\gluon[\E\REG](\pbackx,\pbacky)[9]
\put(\fermionbackx,\fermionbacky){\circle*{500}}
\drawline\fermion[\NE\REG](\fermionbackx,\fermionbacky)[6750]
\drawline\fermion[\NW\REG](\gluonbackx,\gluonbacky)[6750]
\put(\fermionbackx,\fermionbacky){\circle{500}}
\put(\gluonbackx,\gluonbacky){\circle*{500}}
\drawline\fermion[\SE\REG](\gluonbackx,\gluonbacky)[3250]
\end{picture}
\ed
\bd
\begin{picture}(40000,15000)
\put(0,0){3.9}
\drawline\fermion[\NE\REG](0,2000)[3250]
\drawline\gluon[\E\REG](\pbackx,\pbacky)[9]
\drawline\fermion[\NE\REG](\fermionbackx,\fermionbacky)[6750]
\drawline\fermion[\NW\REG](\gluonbackx,\gluonbacky)[6750]
\global\advance\fermionbackx by 300
\global\advance\fermionbacky by -300
\drawline\fermion[\W\REG](\fermionbackx,\fermionbacky)[600]
\drawline\fermion[\N\REG](\fermionbackx,\fermionbacky)[600]
\drawline\fermion[\E\REG](\fermionbackx,\fermionbacky)[600]
\drawline\fermion[\S\REG](\fermionbackx,\fermionbacky)[600]
\drawline\fermion[\SE\REG](\gluonbackx,\gluonbacky)[3250]
\put(20000,0){3.10}
\drawline\fermion[\NE\REG](20000,2000)[3250]
\drawline\gluon[\E\REG](\pbackx,\pbacky)[9]
\put(\fermionbackx,\fermionbacky){\circle*{500}}
\drawline\fermion[\NE\REG](\fermionbackx,\fermionbacky)[6750]
\drawline\fermion[\NW\REG](\gluonbackx,\gluonbacky)[6750]
\global\advance\fermionbackx by 300
\global\advance\fermionbacky by -300
\drawline\fermion[\W\REG](\fermionbackx,\fermionbacky)[600]
\drawline\fermion[\N\REG](\fermionbackx,\fermionbacky)[600]
\drawline\fermion[\E\REG](\fermionbackx,\fermionbacky)[600]
\drawline\fermion[\S\REG](\fermionbackx,\fermionbacky)[600]
\drawline\fermion[\SE\REG](\gluonbackx,\gluonbacky)[3250]
\end{picture}
\ed
\bd
\begin{picture}(40000,15000)
\put(0,0){3.11}
\drawline\fermion[\NE\REG](0,2000)[3250]
\drawline\gluon[\E\REG](\pbackx,\pbacky)[9]
\drawline\fermion[\NE\REG](\fermionbackx,\fermionbacky)[6750]
\drawline\fermion[\NW\REG](\gluonbackx,\gluonbacky)[6750]
\global\advance\fermionbackx by 300
\global\advance\fermionbacky by -300
\drawline\fermion[\W\REG](\fermionbackx,\fermionbacky)[600]
\drawline\fermion[\N\REG](\fermionbackx,\fermionbacky)[600]
\drawline\fermion[\E\REG](\fermionbackx,\fermionbacky)[600]
\drawline\fermion[\S\REG](\fermionbackx,\fermionbacky)[600]
\put(\gluonbackx,\gluonbacky){\circle*{500}}
\drawline\fermion[\SE\REG](\gluonbackx,\gluonbacky)[3250]
\put(20000,0){3.12}
\drawline\fermion[\NE\REG](20000,2000)[3250]
\drawline\gluon[\E\REG](\pbackx,\pbacky)[9]
\put(\fermionbackx,\fermionbacky){\circle*{500}}
\drawline\fermion[\NE\REG](\fermionbackx,\fermionbacky)[6750]
\drawline\fermion[\NW\REG](\gluonbackx,\gluonbacky)[6750]
\global\advance\fermionbackx by 300
\global\advance\fermionbacky by -300
\drawline\fermion[\W\REG](\fermionbackx,\fermionbacky)[600]
\drawline\fermion[\N\REG](\fermionbackx,\fermionbacky)[600]
\drawline\fermion[\E\REG](\fermionbackx,\fermionbacky)[600]
\drawline\fermion[\S\REG](\fermionbackx,\fermionbacky)[600]
\put(\gluonbackx,\gluonbacky){\circle*{500}}
\drawline\fermion[\SE\REG](\gluonbackx,\gluonbacky)[3250]
\end{picture}
\ed
{\small {\bf Fig. 3} - The 12 diagrams that contribute to the graph of Fig.~2.
We have indicated the improved quark-gluon vertex with a dot and the
insertions of $O(a$) and $O(a^2)$ corrections to the vertex operators
with respectively an open circle and a square.}
\vspace{1 cm}

With this understanding we show below only the general patterns of the various
graphs, as representatives of the full set of improved diagrams.

The graphs needed for the 1-loop computation of the matrix element \linebreak
$<q| (O^q_{\mu \nu})^{IMPR} |q>$ and the graphs contributing in the quenched
approximation to $<g,\sigma| O^g_{\mu \nu} |g,\sigma>$
are shown in Figs.~4.1 to 4.8 and Figs.~5.1 to 5.13 respectively,
where the insertion of the vertex operator is always indicated by a dot.
\bd
\begin{picture}(40000,15000)
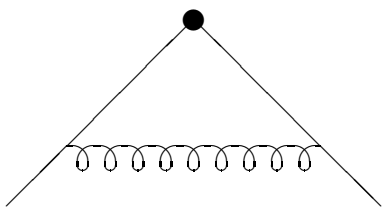
\put(0,0){4.1 - Vertex}
\drawline\fermion[\NE\REG](0,2000)[3250]
\drawline\gluon[\E\REG](\pbackx,\pbacky)[9]
\drawline\fermion[\NE\REG](\fermionbackx,\fermionbacky)[6750]
\put(\fermionbackx,\fermionbacky){\circle*{800}}
\drawline\fermion[\NW\REG](\gluonbackx,\gluonbacky)[6750]
\drawline\fermion[\SE\REG](\gluonbackx,\gluonbacky)[3250]
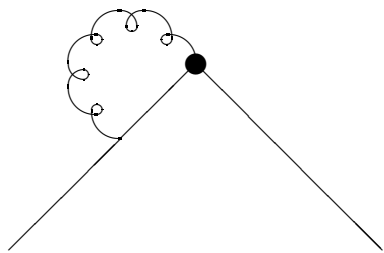
\put(20000,0){4.2 - Sail}
\drawline\fermion[\NE\REG](20000,2000)[6000]
\drawloop\gluon[\NW 5](\pbackx,\pbacky)
\drawline\fermion[\NE\REG](\fermionbackx,\fermionbacky)[4000]
\put(\fermionbackx,\fermionbacky){\circle*{800}}
\drawline\fermion[\SE\REG](\fermionbackx,\fermionbacky)[10000]
\end{picture}
\ed
\bd
\begin{picture}(40000,15000)
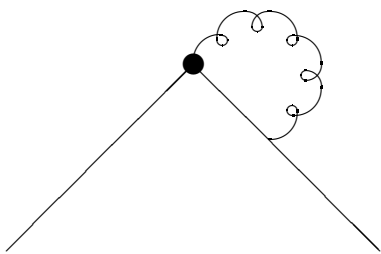
\put(0,0){4.3 - Sail}
\drawline\fermion[\NE\REG](0,2000)[10000]
\put(\fermionbackx,\fermionbacky){\circle*{800}}
\drawloop\gluon[\NE 5](\pbackx,\pbacky)
\drawline\fermion[\SE\REG](\fermionbackx,\fermionbacky)[10000]
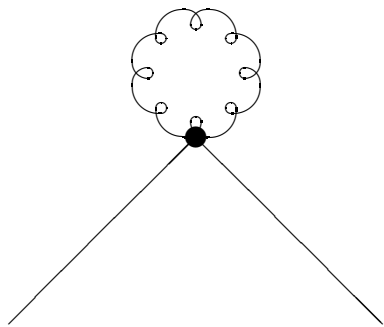
\put(20000,0){4.4 - Operator tadpole}
\drawline\fermion[\NE\REG](20000,2000)[10000]
\put(\fermionbackx,\fermionbacky){\circle*{800}}
\global\advance\pbackx by -400
\drawloop\gluon[\NW 8](\pbackx,\pbacky)
\drawline\fermion[\SE\REG](\fermionbackx,\fermionbacky)[10000]
\end{picture}
\ed
\bd
\begin{picture}(40000,15000)
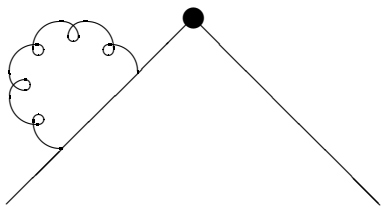
\put(0,0){4.5 - Self-energy}
\drawline\fermion[\NE\REG](0,2000)[3000]
\drawloop\gluon[\NW 5](\pbackx,\pbacky)
\drawline\fermion[\NE\REG](\fermionbackx,\fermionbacky)[7000]
\put(\fermionbackx,\fermionbacky){\circle*{800}}
\drawline\fermion[\SE\REG](\fermionbackx,\fermionbacky)[10000]
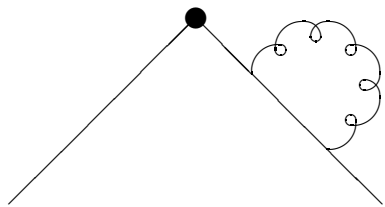
\put(20000,0){4.6 - Self-energy}
\drawline\fermion[\NE\REG](20000,2000)[10000]
\put(\fermionbackx,\fermionbacky){\circle*{800}}
\drawline\fermion[\SE\REG](\fermionbackx,\fermionbacky)[3000]
\drawloop\gluon[\NE 5](\pbackx,\pbacky)
\drawline\fermion[\SE\REG](\fermionbackx,\fermionbacky)[7000]
\end{picture}
\ed
\bd
\begin{picture}(40000,15000)
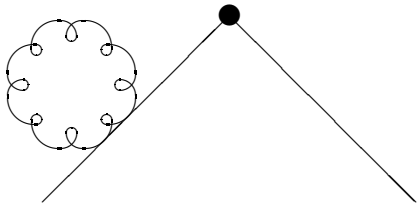
\put(0,0){4.7 - Leg tadpole}
\drawline\fermion[\NE\REG](0,2000)[5000]
\global\advance\pbacky by 500
\drawloop\gluon[\SW 8](\pbackx,\pbacky)
\drawline\fermion[\NE\REG](\fermionbackx,\fermionbacky)[5000]
\put(\fermionbackx,\fermionbacky){\circle*{800}}
\drawline\fermion[\SE\REG](\fermionbackx,\fermionbacky)[10000]
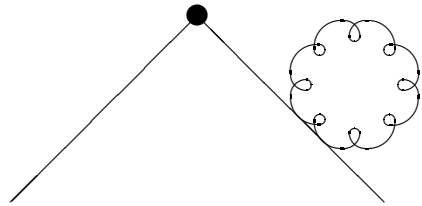
\put(20000,0){4.8 - Leg tadpole}
\drawline\fermion[\NW\REG](35000,2000)[5000]
\global\advance\pbacky by -1500
\global\advance\pbackx by 2000
\drawloop\gluon[\NW 8](\pbackx,\pbacky)
\drawline\fermion[\NW\REG](\fermionbackx,\fermionbacky)[5000]
\put(\fermionbackx,\fermionbacky){\circle*{800}}
\drawline\fermion[\SW\REG](\fermionbackx,\fermionbacky)[10000]
\end{picture}
\ed
{\small {\bf Fig. 4} - The different types of graphs contributing to the 1-loop
approximation of the matrix element $<q| (O^q_{\mu \nu})^{IMPR} |q>$.}

\bd
\begin{picture}(40000,15000)
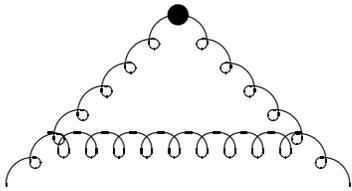
\put(0,0){5.1 - Vertex}
\drawline\gluon[\NE \REG](0,2000)[6]
\put(\gluonbackx,\gluonbacky){\circle*{800}}
\drawline\gluon[\SE \REG](\pbackx,\pbacky)[6]
\global\advance\pmidx by 1400
\global\advance\pmidy by -1200
\drawline\gluon[\W \FLIPPED](\pmidx,\pmidy)[9]
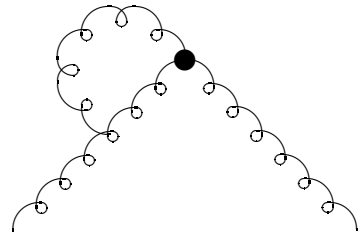
\put(20000,0){5.2 - Sail}
\drawline\gluon[\NW \FLIPPED](35000,2000)[6]
\put(\gluonbackx,\gluonbacky){\circle*{800}}
\drawline\gluon[\SW \FLIPPED](\pbackx,\pbacky)[6]
\divide\unitboxwidth by 2
\divide\unitboxheight by 2
\global\advance\pmidx by -\unitboxwidth
\global\advance\pmidy by -\unitboxheight
\drawloop\gluon[\NW 5](\pmidx,\pmidy)
\end{picture}
\ed
\bd
\begin{picture}(40000,15000)
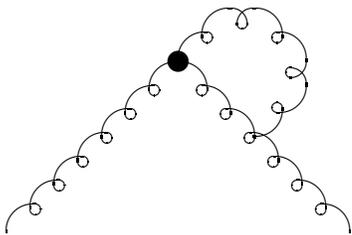
\put(0,0){5.3 - Sail}
\drawline\gluon[\NE \REG](0,2000)[6]
\put(\gluonbackx,\gluonbacky){\circle*{800}}
\drawline\gluon[\SE \REG](\pbackx,\pbacky)[6]
\drawloop\gluon[\NE 5](\pfrontx,\pfronty)
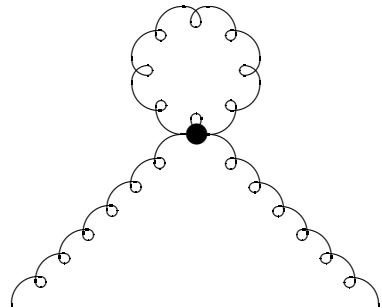
\put(20000,0){5.4 - Operator tadpole}
\drawline\gluon[\NE \REG](20000,2000)[6]
\global\advance\gluonbackx by 500
\put(\gluonbackx,\gluonbacky){\circle*{800}}
\global\advance\gluonbackx by 400
\drawloop\gluon[\W 8](\gluonbackx,\gluonbacky)
\drawline\gluon[\SE \REG](\pbackx,\pbacky)[6]
\end{picture}
\ed
\bd
\begin{picture}(40000,15000)
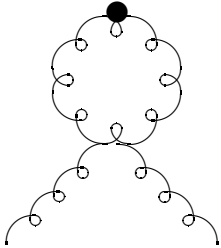
\put(0,0){5.5 - Tadpole (QCD vertex)}
\drawline\gluon[\NE \REG](0,2000)[3]
\global\advance\pbackx by 800
\drawloop\gluon[\W 8](\pbackx,\pbacky)
\global\advance\pbackx by -400
\global\advance\pbacky by 5000
\put(\pbackx,\pbacky){\circle*{800}}
\global\advance\pbacky by -5000
\drawline\gluon[\SE \REG](\pbackx,\pbacky)[3]
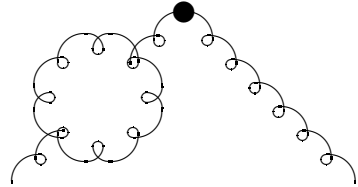
\put(20000,0){5.6 - Gluon loop }
\startphantom
\drawline\gluon[\NE \REG](20000,2000)[6]
\stopphantom
\put(\gluonbackx,\gluonbacky){\circle*{800}}
\drawloop\gluon[\NE 0](\pmidx,\pmidy)
\drawline\gluon[\NE \REG](20000,2000)[1]
\startphantom
\drawline\gluon[\NE \REG](20000,2000)[6]
\stopphantom
\drawline\gluon[\SW \FLIPPED](\pbackx,\pbacky)[1]
\drawline\gluon[\SE \REG](\pfrontx,\pfronty)[6]
\end{picture}
\ed
\bd
\begin{picture}(40000,15000)
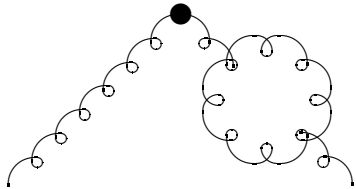
\put(0,0){5.7 - Gluon loop}
\startphantom
\drawline\gluon[\NW \FLIPPED](15000,2000)[6]
\stopphantom
\put(\gluonbackx,\gluonbacky){\circle*{800}}
\drawloop\gluon[\NW 0](\pmidx,\pmidy)
\drawline\gluon[\NW \FLIPPED](15000,2000)[1]
\startphantom
\drawline\gluon[\NW \FLIPPED](15000,2000)[6]
\stopphantom
\drawline\gluon[\SE \REG](\pbackx,\pbacky)[1]
\drawline\gluon[\SW \FLIPPED](\pfrontx,\pfronty)[6]
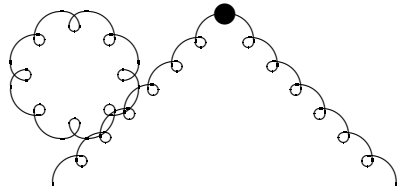
\put(20000,0){5.8 - Leg tadpole}
\drawline\gluon[\NE \REG](20000,2000)[6]
\put(\gluonbackx,\gluonbacky){\circle*{800}}
\global\advance\pmidy by 500
\drawloop\gluon[\SW 8](\pmidx,\pmidy)
\startphantom
\drawline\gluon[\NE \REG](20000,2000)[6]
\stopphantom
\drawline\gluon[\SE \REG](\gluonbackx,\gluonbacky)[6]
\end{picture}
\ed
\bd
\begin{picture}(40000,15000)
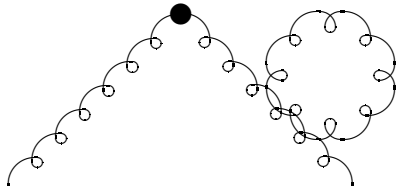
\put(0,0){5.9 - Leg tadpole}
\drawline\gluon[\NW \FLIPPED](15000,2000)[6]
\put(\gluonbackx,\gluonbacky){\circle*{800}}
\global\advance\pmidy by -1500
\global\advance\pmidx by 2000
\drawloop\gluon[\NW 8](\pmidx,\pmidy)
\startphantom
\drawline\gluon[\NW \FLIPPED](15000,2000)[6]
\stopphantom
\drawline\gluon[\SW \FLIPPED](\pbackx,\pbacky)[6]
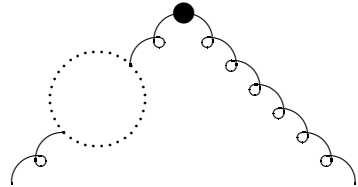
\put(20000,0){5.10 - Ghost}
\startphantom
\drawline\gluon[\NE \REG](20000,2000)[6]
\stopphantom
\put(\gluonbackx,\gluonbacky){\circle*{800}}
\global\Xone=\pmidx
\global\Yone=\pmidy
\global\advance\pmidx by  1793
\global\advance\pmidy by  157
\put(\pmidx,\pmidy){\circle*{1}}
\global\advance\pmidx by  -3586
\put(\pmidx,\pmidy){\circle*{1}}
\global\advance\pmidy by  -314
\put(\pmidx,\pmidy){\circle*{1}}
\global\advance\pmidx by  3586
\put(\pmidx,\pmidy){\circle*{1}}
\global\pmidx=\Xone
\global\pmidy=\Yone
\global\advance\pmidx by  1739
\global\advance\pmidy by  466
\put(\pmidx,\pmidy){\circle*{1}}
\global\advance\pmidx by  -3478
\put(\pmidx,\pmidy){\circle*{1}}
\global\advance\pmidy by  -932
\put(\pmidx,\pmidy){\circle*{1}}
\global\advance\pmidx by  3478
\put(\pmidx,\pmidy){\circle*{1}}
\global\pmidx=\Xone
\global\pmidy=\Yone
\global\advance\pmidx by  1631
\global\advance\pmidy by  761
\put(\pmidx,\pmidy){\circle*{1}}
\global\advance\pmidx by  -3262
\put(\pmidx,\pmidy){\circle*{1}}
\global\advance\pmidy by  -1522
\put(\pmidx,\pmidy){\circle*{1}}
\global\advance\pmidx by  3262
\put(\pmidx,\pmidy){\circle*{1}}
\global\pmidx=\Xone
\global\pmidy=\Yone
\global\advance\pmidx by  1474
\global\advance\pmidy by  1032
\put(\pmidx,\pmidy){\circle*{1}}
\global\advance\pmidx by  -2948
\put(\pmidx,\pmidy){\circle*{1}}
\global\advance\pmidy by  -2064
\put(\pmidx,\pmidy){\circle*{1}}
\global\advance\pmidx by  2948
\put(\pmidx,\pmidy){\circle*{1}}
\global\pmidx=\Xone
\global\pmidy=\Yone
\global\advance\pmidx by  1273
\global\advance\pmidy by  1273
\put(\pmidx,\pmidy){\circle*{1}}
\global\advance\pmidx by  -2546
\put(\pmidx,\pmidy){\circle*{1}}
\global\advance\pmidy by  -2546
\put(\pmidx,\pmidy){\circle*{1}}
\global\advance\pmidx by  2546
\put(\pmidx,\pmidy){\circle*{1}}
\global\pmidx=\Xone
\global\pmidy=\Yone
\global\advance\pmidx by  1032
\global\advance\pmidy by  1474
\put(\pmidx,\pmidy){\circle*{1}}
\global\advance\pmidx by  -2064
\put(\pmidx,\pmidy){\circle*{1}}
\global\advance\pmidy by  -2948
\put(\pmidx,\pmidy){\circle*{1}}
\global\advance\pmidx by  2064
\put(\pmidx,\pmidy){\circle*{1}}
\global\pmidx=\Xone
\global\pmidy=\Yone
\global\advance\pmidx by  761
\global\advance\pmidy by  1631
\put(\pmidx,\pmidy){\circle*{1}}
\global\advance\pmidx by  -1522
\put(\pmidx,\pmidy){\circle*{1}}
\global\advance\pmidy by  -3262
\put(\pmidx,\pmidy){\circle*{1}}
\global\advance\pmidx by  1522
\put(\pmidx,\pmidy){\circle*{1}}
\global\pmidx=\Xone
\global\pmidy=\Yone
\global\advance\pmidx by  466
\global\advance\pmidy by  1739
\put(\pmidx,\pmidy){\circle*{1}}
\global\advance\pmidx by  -932
\put(\pmidx,\pmidy){\circle*{1}}
\global\advance\pmidy by  -3478
\put(\pmidx,\pmidy){\circle*{1}}
\global\advance\pmidx by  932
\put(\pmidx,\pmidy){\circle*{1}}
\global\pmidx=\Xone
\global\pmidy=\Yone
\global\advance\pmidx by  157
\global\advance\pmidy by  1793
\put(\pmidx,\pmidy){\circle*{1}}
\global\advance\pmidx by  -314
\put(\pmidx,\pmidy){\circle*{1}}
\global\advance\pmidy by  -3586
\put(\pmidx,\pmidy){\circle*{1}}
\global\advance\pmidx by  314
\put(\pmidx,\pmidy){\circle*{1}}
\global\pmidx=\Xone
\global\pmidy=\Yone
\drawline\gluon[\NE \REG](20000,2000)[1]
\startphantom
\drawline\gluon[\NE \REG](20000,2000)[6]
\stopphantom
\drawline\gluon[\SW \FLIPPED](\pbackx,\pbacky)[1]
\drawline\gluon[\SE \REG](\pfrontx,\pfronty)[6]
\end{picture}
\ed
\bd
\begin{picture}(40000,15000)
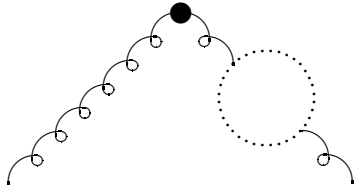
\put(0,0){5.11 - Ghost}
\startphantom
\drawline\gluon[\NW \FLIPPED](15000,2000)[6]
\stopphantom
\put(\gluonbackx,\gluonbacky){\circle*{800}}
\global\Xone=\pmidx
\global\Yone=\pmidy
\global\advance\pmidx by  1793
\global\advance\pmidy by  157
\put(\pmidx,\pmidy){\circle*{1}}
\global\advance\pmidx by  -3586
\put(\pmidx,\pmidy){\circle*{1}}
\global\advance\pmidy by  -314
\put(\pmidx,\pmidy){\circle*{1}}
\global\advance\pmidx by  3586
\put(\pmidx,\pmidy){\circle*{1}}
\global\pmidx=\Xone
\global\pmidy=\Yone
\global\advance\pmidx by  1739
\global\advance\pmidy by  466
\put(\pmidx,\pmidy){\circle*{1}}
\global\advance\pmidx by  -3478
\put(\pmidx,\pmidy){\circle*{1}}
\global\advance\pmidy by  -932
\put(\pmidx,\pmidy){\circle*{1}}
\global\advance\pmidx by  3478
\put(\pmidx,\pmidy){\circle*{1}}
\global\pmidx=\Xone
\global\pmidy=\Yone
\global\advance\pmidx by  1631
\global\advance\pmidy by  761
\put(\pmidx,\pmidy){\circle*{1}}
\global\advance\pmidx by  -3262
\put(\pmidx,\pmidy){\circle*{1}}
\global\advance\pmidy by  -1522
\put(\pmidx,\pmidy){\circle*{1}}
\global\advance\pmidx by  3262
\put(\pmidx,\pmidy){\circle*{1}}
\global\pmidx=\Xone
\global\pmidy=\Yone
\global\advance\pmidx by  1474
\global\advance\pmidy by  1032
\put(\pmidx,\pmidy){\circle*{1}}
\global\advance\pmidx by  -2948
\put(\pmidx,\pmidy){\circle*{1}}
\global\advance\pmidy by  -2064
\put(\pmidx,\pmidy){\circle*{1}}
\global\advance\pmidx by  2948
\put(\pmidx,\pmidy){\circle*{1}}
\global\pmidx=\Xone
\global\pmidy=\Yone
\global\advance\pmidx by  1273
\global\advance\pmidy by  1273
\put(\pmidx,\pmidy){\circle*{1}}
\global\advance\pmidx by  -2546
\put(\pmidx,\pmidy){\circle*{1}}
\global\advance\pmidy by  -2546
\put(\pmidx,\pmidy){\circle*{1}}
\global\advance\pmidx by  2546
\put(\pmidx,\pmidy){\circle*{1}}
\global\pmidx=\Xone
\global\pmidy=\Yone
\global\advance\pmidx by  1032
\global\advance\pmidy by  1474
\put(\pmidx,\pmidy){\circle*{1}}
\global\advance\pmidx by  -2064
\put(\pmidx,\pmidy){\circle*{1}}
\global\advance\pmidy by  -2948
\put(\pmidx,\pmidy){\circle*{1}}
\global\advance\pmidx by  2064
\put(\pmidx,\pmidy){\circle*{1}}
\global\pmidx=\Xone
\global\pmidy=\Yone
\global\advance\pmidx by  761
\global\advance\pmidy by  1631
\put(\pmidx,\pmidy){\circle*{1}}
\global\advance\pmidx by  -1522
\put(\pmidx,\pmidy){\circle*{1}}
\global\advance\pmidy by  -3262
\put(\pmidx,\pmidy){\circle*{1}}
\global\advance\pmidx by  1522
\put(\pmidx,\pmidy){\circle*{1}}
\global\pmidx=\Xone
\global\pmidy=\Yone
\global\advance\pmidx by  466
\global\advance\pmidy by  1739
\put(\pmidx,\pmidy){\circle*{1}}
\global\advance\pmidx by  -932
\put(\pmidx,\pmidy){\circle*{1}}
\global\advance\pmidy by  -3478
\put(\pmidx,\pmidy){\circle*{1}}
\global\advance\pmidx by  932
\put(\pmidx,\pmidy){\circle*{1}}
\global\pmidx=\Xone
\global\pmidy=\Yone
\global\advance\pmidx by  157
\global\advance\pmidy by  1793
\put(\pmidx,\pmidy){\circle*{1}}
\global\advance\pmidx by  -314
\put(\pmidx,\pmidy){\circle*{1}}
\global\advance\pmidy by  -3586
\put(\pmidx,\pmidy){\circle*{1}}
\global\advance\pmidx by  314
\put(\pmidx,\pmidy){\circle*{1}}
\global\pmidx=\Xone
\global\pmidy=\Yone
\drawline\gluon[\NW \FLIPPED](15000,2000)[1]
\startphantom
\drawline\gluon[\NW \FLIPPED](15000,2000)[6]
\stopphantom
\drawline\gluon[\SE \REG](\pbackx,\pbacky)[1]
\drawline\gluon[\SW \FLIPPED](\pfrontx,\pfronty)[6]
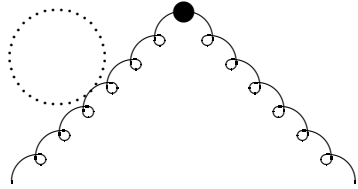
\put(20000,0){5.12 - Ghost}
\drawline\gluon[\NE \REG](20000,2000)[6]
\put(\gluonbackx,\gluonbacky){\circle*{800}}
\divide\unitboxwidth by 10
\divide\unitboxheight by 10
\multiply\unitboxwidth by 17
\multiply\unitboxheight by 17
\global\advance\pmidx by -\unitboxwidth
\global\advance\pmidy by \unitboxheight
\global\Xone=\pmidx
\global\Yone=\pmidy
\global\advance\pmidx by  1793
\global\advance\pmidy by  157
\put(\pmidx,\pmidy){\circle*{1}}
\global\advance\pmidx by  -3586
\put(\pmidx,\pmidy){\circle*{1}}
\global\advance\pmidy by  -314
\put(\pmidx,\pmidy){\circle*{1}}
\global\advance\pmidx by  3586
\put(\pmidx,\pmidy){\circle*{1}}
\global\pmidx=\Xone
\global\pmidy=\Yone
\global\advance\pmidx by  1739
\global\advance\pmidy by  466
\put(\pmidx,\pmidy){\circle*{1}}
\global\advance\pmidx by  -3478
\put(\pmidx,\pmidy){\circle*{1}}
\global\advance\pmidy by  -932
\put(\pmidx,\pmidy){\circle*{1}}
\global\advance\pmidx by  3478
\put(\pmidx,\pmidy){\circle*{1}}
\global\pmidx=\Xone
\global\pmidy=\Yone
\global\advance\pmidx by  1631
\global\advance\pmidy by  761
\put(\pmidx,\pmidy){\circle*{1}}
\global\advance\pmidx by  -3262
\put(\pmidx,\pmidy){\circle*{1}}
\global\advance\pmidy by  -1522
\put(\pmidx,\pmidy){\circle*{1}}
\global\advance\pmidx by  3262
\put(\pmidx,\pmidy){\circle*{1}}
\global\pmidx=\Xone
\global\pmidy=\Yone
\global\advance\pmidx by  1474
\global\advance\pmidy by  1032
\put(\pmidx,\pmidy){\circle*{1}}
\global\advance\pmidx by  -2948
\put(\pmidx,\pmidy){\circle*{1}}
\global\advance\pmidy by  -2064
\put(\pmidx,\pmidy){\circle*{1}}
\global\advance\pmidx by  2948
\put(\pmidx,\pmidy){\circle*{1}}
\global\pmidx=\Xone
\global\pmidy=\Yone
\global\advance\pmidx by  1273
\global\advance\pmidy by  1273
\put(\pmidx,\pmidy){\circle*{1}}
\global\advance\pmidx by  -2546
\put(\pmidx,\pmidy){\circle*{1}}
\global\advance\pmidy by  -2546
\put(\pmidx,\pmidy){\circle*{1}}
\global\advance\pmidx by  2546
\put(\pmidx,\pmidy){\circle*{1}}
\global\pmidx=\Xone
\global\pmidy=\Yone
\global\advance\pmidx by  1032
\global\advance\pmidy by  1474
\put(\pmidx,\pmidy){\circle*{1}}
\global\advance\pmidx by  -2064
\put(\pmidx,\pmidy){\circle*{1}}
\global\advance\pmidy by  -2948
\put(\pmidx,\pmidy){\circle*{1}}
\global\advance\pmidx by  2064
\put(\pmidx,\pmidy){\circle*{1}}
\global\pmidx=\Xone
\global\pmidy=\Yone
\global\advance\pmidx by  761
\global\advance\pmidy by  1631
\put(\pmidx,\pmidy){\circle*{1}}
\global\advance\pmidx by  -1522
\put(\pmidx,\pmidy){\circle*{1}}
\global\advance\pmidy by  -3262
\put(\pmidx,\pmidy){\circle*{1}}
\global\advance\pmidx by  1522
\put(\pmidx,\pmidy){\circle*{1}}
\global\pmidx=\Xone
\global\pmidy=\Yone
\global\advance\pmidx by  466
\global\advance\pmidy by  1739
\put(\pmidx,\pmidy){\circle*{1}}
\global\advance\pmidx by  -932
\put(\pmidx,\pmidy){\circle*{1}}
\global\advance\pmidy by  -3478
\put(\pmidx,\pmidy){\circle*{1}}
\global\advance\pmidx by  932
\put(\pmidx,\pmidy){\circle*{1}}
\global\pmidx=\Xone
\global\pmidy=\Yone
\global\advance\pmidx by  157
\global\advance\pmidy by  1793
\put(\pmidx,\pmidy){\circle*{1}}
\global\advance\pmidx by  -314
\put(\pmidx,\pmidy){\circle*{1}}
\global\advance\pmidy by  -3586
\put(\pmidx,\pmidy){\circle*{1}}
\global\advance\pmidx by  314
\put(\pmidx,\pmidy){\circle*{1}}
\global\pmidx=\Xone
\global\pmidy=\Yone
\startphantom
\drawline\gluon[\NE \REG](20000,2000)[6]
\stopphantom
\drawline\gluon[\SE \REG](\gluonbackx,\gluonbacky)[6]
\end{picture}
\ed
\bd
\begin{picture}(40000,15000)
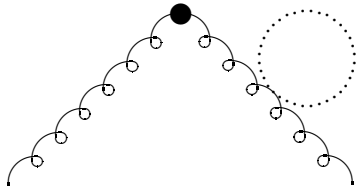
\put(0,0){5.13 - Ghost}
\drawline\gluon[\NW \FLIPPED](15000,2000)[6]
\put(\gluonbackx,\gluonbacky){\circle*{800}}
\divide\unitboxwidth by 10
\divide\unitboxheight by 10
\multiply\unitboxwidth by 17
\multiply\unitboxheight by 17
\global\advance\pmidx by -\unitboxwidth
\global\advance\pmidy by \unitboxheight
\global\Xone=\pmidx
\global\Yone=\pmidy
\global\advance\pmidx by  1793
\global\advance\pmidy by  157
\put(\pmidx,\pmidy){\circle*{1}}
\global\advance\pmidx by  -3586
\put(\pmidx,\pmidy){\circle*{1}}
\global\advance\pmidy by  -314
\put(\pmidx,\pmidy){\circle*{1}}
\global\advance\pmidx by  3586
\put(\pmidx,\pmidy){\circle*{1}}
\global\pmidx=\Xone
\global\pmidy=\Yone
\global\advance\pmidx by  1739
\global\advance\pmidy by  466
\put(\pmidx,\pmidy){\circle*{1}}
\global\advance\pmidx by  -3478
\put(\pmidx,\pmidy){\circle*{1}}
\global\advance\pmidy by  -932
\put(\pmidx,\pmidy){\circle*{1}}
\global\advance\pmidx by  3478
\put(\pmidx,\pmidy){\circle*{1}}
\global\pmidx=\Xone
\global\pmidy=\Yone
\global\advance\pmidx by  1631
\global\advance\pmidy by  761
\put(\pmidx,\pmidy){\circle*{1}}
\global\advance\pmidx by  -3262
\put(\pmidx,\pmidy){\circle*{1}}
\global\advance\pmidy by  -1522
\put(\pmidx,\pmidy){\circle*{1}}
\global\advance\pmidx by  3262
\put(\pmidx,\pmidy){\circle*{1}}
\global\pmidx=\Xone
\global\pmidy=\Yone
\global\advance\pmidx by  1474
\global\advance\pmidy by  1032
\put(\pmidx,\pmidy){\circle*{1}}
\global\advance\pmidx by  -2948
\put(\pmidx,\pmidy){\circle*{1}}
\global\advance\pmidy by  -2064
\put(\pmidx,\pmidy){\circle*{1}}
\global\advance\pmidx by  2948
\put(\pmidx,\pmidy){\circle*{1}}
\global\pmidx=\Xone
\global\pmidy=\Yone
\global\advance\pmidx by  1273
\global\advance\pmidy by  1273
\put(\pmidx,\pmidy){\circle*{1}}
\global\advance\pmidx by  -2546
\put(\pmidx,\pmidy){\circle*{1}}
\global\advance\pmidy by  -2546
\put(\pmidx,\pmidy){\circle*{1}}
\global\advance\pmidx by  2546
\put(\pmidx,\pmidy){\circle*{1}}
\global\pmidx=\Xone
\global\pmidy=\Yone
\global\advance\pmidx by  1032
\global\advance\pmidy by  1474
\put(\pmidx,\pmidy){\circle*{1}}
\global\advance\pmidx by  -2064
\put(\pmidx,\pmidy){\circle*{1}}
\global\advance\pmidy by  -2948
\put(\pmidx,\pmidy){\circle*{1}}
\global\advance\pmidx by  2064
\put(\pmidx,\pmidy){\circle*{1}}
\global\pmidx=\Xone
\global\pmidy=\Yone
\global\advance\pmidx by  761
\global\advance\pmidy by  1631
\put(\pmidx,\pmidy){\circle*{1}}
\global\advance\pmidx by  -1522
\put(\pmidx,\pmidy){\circle*{1}}
\global\advance\pmidy by  -3262
\put(\pmidx,\pmidy){\circle*{1}}
\global\advance\pmidx by  1522
\put(\pmidx,\pmidy){\circle*{1}}
\global\pmidx=\Xone
\global\pmidy=\Yone
\global\advance\pmidx by  466
\global\advance\pmidy by  1739
\put(\pmidx,\pmidy){\circle*{1}}
\global\advance\pmidx by  -932
\put(\pmidx,\pmidy){\circle*{1}}
\global\advance\pmidy by  -3478
\put(\pmidx,\pmidy){\circle*{1}}
\global\advance\pmidx by  932
\put(\pmidx,\pmidy){\circle*{1}}
\global\pmidx=\Xone
\global\pmidy=\Yone
\global\advance\pmidx by  157
\global\advance\pmidy by  1793
\put(\pmidx,\pmidy){\circle*{1}}
\global\advance\pmidx by  -314
\put(\pmidx,\pmidy){\circle*{1}}
\global\advance\pmidy by  -3586
\put(\pmidx,\pmidy){\circle*{1}}
\global\advance\pmidx by  314
\put(\pmidx,\pmidy){\circle*{1}}
\global\pmidx=\Xone
\global\pmidy=\Yone
\startphantom
\drawline\gluon[\NW \FLIPPED](15000,2000)[6]
\stopphantom
\drawline\gluon[\SW \FLIPPED](\pbackx,\pbacky)[6]
\end{picture}
\ed
{\small {\bf Fig. 5} - The different types of graphs contributing to the 1-loop
approximation of the matrix element $<g,\sigma| O^g_{\mu \nu} |g,\sigma>$,
in the quenched approximation. The dotted line represents a ghost loop.}

\bd
\begin{picture}(40000,15000)
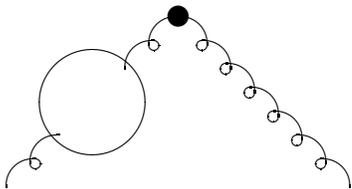
\put(0,0){6.1 - Quark loop}
\startphantom
\drawline\gluon[\NE \REG](0,2000)[6]
\stopphantom
\put(\gluonbackx,\gluonbacky){\circle*{800}}
\put(\pmidx,\pmidy){\circle{6000}}
\drawline\gluon[\NE \REG](0,2000)[1]
\startphantom
\drawline\gluon[\NE \REG](0,2000)[6]
\stopphantom
\drawline\gluon[\SW \FLIPPED](\pbackx,\pbacky)[1]
\drawline\gluon[\SE \REG](\pfrontx,\pfronty)[6]
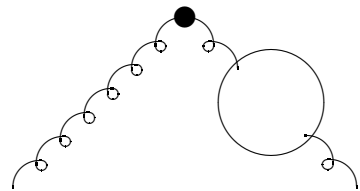
\put(20000,0){6.2 - Quark loop}
\startphantom
\drawline\gluon[\NW \FLIPPED](35000,2000)[6]
\stopphantom
\put(\gluonbackx,\gluonbacky){\circle*{800}}
\put(\pmidx,\pmidy){\circle{6000}}
\drawline\gluon[\NW \FLIPPED](35000,2000)[1]
\startphantom
\drawline\gluon[\NW \FLIPPED](35000,2000)[6]
\stopphantom
\drawline\gluon[\SE \REG](\pbackx,\pbacky)[1]
\drawline\gluon[\SW \FLIPPED](\pfrontx,\pfronty)[6]
\end{picture}
\ed
\bd
\begin{picture}(40000,15000)
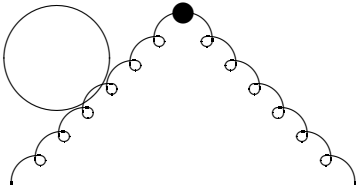
\put(0,0){6.3 - Tadpole-quark loop}
\drawline\gluon[\NE \REG](0,2000)[6]
\put(\gluonbackx,\gluonbacky){\circle*{800}}
\divide\unitboxwidth by 10
\divide\unitboxheight by 10
\multiply\unitboxwidth by 17
\multiply\unitboxheight by 17
\global\advance\pmidx by -\unitboxwidth
\global\advance\pmidy by \unitboxheight
\put(\pmidx,\pmidy){\circle{6000}}
\startphantom
\drawline\gluon[\NE \REG](0,2000)[6]
\stopphantom
\drawline\gluon[\SE \REG](\gluonbackx,\gluonbacky)[6]
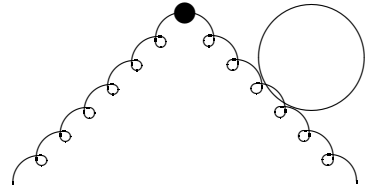
\put(20000,0){6.4 - Tadpole-quark loop}
\drawline\gluon[\NW \FLIPPED](35000,2000)[6]
\put(\gluonbackx,\gluonbacky){\circle*{800}}
\divide\unitboxwidth by 10
\divide\unitboxheight by 10
\multiply\unitboxwidth by 17
\multiply\unitboxheight by 17
\global\advance\pmidx by -\unitboxwidth
\global\advance\pmidy by \unitboxheight
\put(\pmidx,\pmidy){\circle{6000}}
\startphantom
\drawline\gluon[\NW \FLIPPED](35000,2000)[6]
\stopphantom
\drawline\gluon[\SW \FLIPPED](\pbackx,\pbacky)[6]
\end{picture}
\ed
{\small {\bf Fig. 6} - The different types of graphs contributing to the 1-loop
approximation of the matrix element $<g,\sigma| O^g_{\mu \nu} |g,\sigma>$
to be added in the full (unquenched) theory to the diagrams of Figs.~5.}
\vspace{1 cm}

In the full (unquenched) theory there will be mixing between the flavor Singlet
operators $(O^q_{\mu \nu})^{IMPR}$ and $O^g_{\mu \nu}$. To compute the mixing
coefficients we have

1) to add to the diagrams of Figs.~5 those of Figs.~6, in which a quark loop
is present.

2) to evaluate the off-diagonal matrix elements
$<g,\sigma| (O^q_{\mu \nu})^{IMPR} |g,\sigma>$ and
$<q| O^g_{\mu \nu} |q>$.
The corresponding diagrams are listed in Figs.~7 and Figs.~8 respectively.

\bd
\begin{picture}(40000,15000)
\put(0,0){7.1}
\drawline\gluon[\NE\REG](0,2000)[2]
\drawline\fermion[\NE\REG](\pbackx,\pbacky)[6750]
\put(\fermionbackx,\fermionbacky){\circle*{800}}
\drawline\fermion[\E\REG](\pfrontx,\pfronty)[9546]
\drawline\fermion[\NW\REG](\pbackx,\pbacky)[6750]
\drawline\gluon[\SE\REG](\pfrontx,\pfronty)[2]
\put(20000,0){7.2}
\drawline\gluon[\NE \REG](20000,2000)[6]
\put(\gluonbackx,\gluonbacky){\circle*{800}}
\global\advance\pbacky by 2000
\put(\pbackx,\pbacky){\circle{6000}}
\global\advance\pbacky by -2000
\drawline\gluon[\SE \REG](\pbackx,\pbacky)[6]
\end{picture}
\ed
\bd
\begin{picture}(40000,15000)
\put(0,0){7.3}
\drawline\gluon[\NE \REG](0,2000)[3]
\global\advance\pbacky by 2000
\put(\pbackx,\pbacky){\circle{6000}}
\global\advance\pbacky by 2000
\put(\pbackx,\pbacky){\circle*{800}}
\global\advance\pbacky by -4000
\drawline\gluon[\SE \REG](\pbackx,\pbacky)[3]
\put(20000,0){7.4}
\drawline\gluon[\NE \REG](20000,2000)[6]
\put(\gluonbackx,\gluonbacky){\circle*{800}}
\divide\unitboxwidth by 10
\divide\unitboxheight by 10
\multiply\unitboxwidth by 17
\multiply\unitboxheight by 17
\global\advance\pbackx by \unitboxwidth
\global\advance\pbacky by -\unitboxheight
\put(\pbackx,\pbacky){\circle{6000}}
\global\advance\pbackx by \unitboxwidth
\global\advance\pbacky by -\unitboxheight
\global\advance\pbackx by 500
\global\advance\pbacky by 500
\drawline\gluon[\SE \REG](\pbackx,\pbacky)[3]
\end{picture}
\ed
\bd
\begin{picture}(40000,15000)
\put(0,0){7.5}
\drawline\gluon[\NW \FLIPPED](15000,2000)[6]
\put(\gluonbackx,\gluonbacky){\circle*{800}}
\divide\unitboxwidth by 10
\divide\unitboxheight by 10
\multiply\unitboxwidth by 17
\multiply\unitboxheight by 17
\global\advance\pbackx by \unitboxwidth
\global\advance\pbacky by -\unitboxheight
\put(\pbackx,\pbacky){\circle{6000}}
\global\advance\pbackx by \unitboxwidth
\global\advance\pbacky by -\unitboxheight
\global\advance\pbackx by -500
\global\advance\pbacky by 500
\drawline\gluon[\SW \FLIPPED](\pbackx,\pbacky)[3]
\end{picture}
\ed
{\small {\bf Fig. 7} - The different types of graphs contributing to the 1-loop
approximation of the matrix element
$<g,\sigma| (O^q_{\mu \nu})^{IMPR} |g,\sigma>$.}

\bd
\begin{picture}(40000,15000)
\put(0,0){8.1}
\drawline\fermion[\NE\REG](0,2000)[3250]
\drawline\gluon[\NE \REG](\pbackx,\pbacky)[4]
\put(\gluonbackx,\gluonbacky){\circle*{800}}
\drawline\gluon[\SE \REG](\pbackx,\pbacky)[4]
\drawline\fermion[\W\REG](\pbackx,\pbacky)[9546]
\drawline\fermion[\SE\REG](\pfrontx,\pfronty)[3250]
\put(20000,0){8.2}
\drawline\fermion[\NE\REG](20000,2000)[3250]
\drawline\fermion[\SE\REG](\pbackx,\pbacky)[3250]
\global\advance\pfronty by 5000
\put(\pfrontx,\pfronty){\circle*{800}}
\global\advance\pfronty by -2500
\drawloop\gluon[\NW 0](\pfrontx,\pfronty)
\end{picture}
\ed
{\small {\bf Fig. 8} - The different types of graphs contributing to the 1-loop
approximation of the matrix element $<q| O^g_{\mu \nu} |q>$.}

\section*{Appendix C}

\section*{Products of sines}

At some intermediate stage of the computation there appear expressions whose
general structure is of the kind
\be
{\cal I}(\mu_1,\ldots,\mu_{2n}) =
\int\! d^{4} k f(\cos k, \sum_{\lambda} \sin^2 k_{\lambda})
\prod_{i=1}^{2n} \sin k_{\mu_{i}} \label{eq:prodotto},
\ee
By exploiting the H(4) covariance properties of the integrand, the integral
(\ref{eq:prodotto}) is first reduced to a sum of simpler terms that have the
form
\be
\int\! d^{4} k f(\cos k, \sum_{\lambda} \sin^2 k_{\lambda})
\prod_{\mu=1}^{4} \sin^{2 n_{\mu}} k_{\mu} \label{eq:contrazioni},
\ee
where $n_{\mu}$ are integers ranging between 0 and $n$ and satisfying
\be
\sum_{\mu=1}^{4} n_{\mu} = n.
\ee
For example in the $n = 2$ case one gets
\bd
{\cal I}(\mu_1,\mu_2,\mu_3,\mu_4) =
\int\! d^{4} k \, f(\cos k, \sum_{\lambda} \sin^2 k_{\lambda})
\sin k_{\mu_1} \sin k_{\mu_2} \sin k_{\mu_3}
\sin k_{\mu_4} =
\ed
\bd
\int \! d^{4} k \, f(\cos k, \sum_{\lambda} \sin^2 k_{\lambda})
\sin^2 k_{\mu_1} \sin^2 k_{\mu_3}
\: \delta_{\mu_1 \mu_2} \: \delta_{\mu_3 \mu_4}
\Big|_{\mu_1 \neq \mu_3}
\ed
\bd
+\int \! d^{4} k \, f(\cos k, \sum_{\lambda} \sin^2 k_{\lambda})
\sin^2 k_{\mu_1} \sin^2 k_{\mu_2}
\: \delta_{\mu_1 \mu_3} \: \delta_{\mu_2 \mu_4}
\Big|_{\mu_1 \neq \mu_2}
\ed
\bd
+\int \! d^{4} k \, f(\cos k, \sum_{\lambda} \sin^2 k_{\lambda})
\sin^2 k_{\mu_1} \sin^2 k_{\mu_2}
\: \delta_{\mu_1 \mu_4} \: \delta_{\mu_2 \mu_3}
\Big|_{\mu_1 \neq \mu_2}
\ed
\be
+\int\! d^{4} k\, f(\cos k, \sum_{\lambda} \sin^2 k_{\lambda})
\sin^4 k_{\mu_1}
\: \delta_{\mu_1 \mu_2 \mu_3 \mu_4} \label{eq:diversi},
\ee
where $\delta_{\mu_1 \mu_2 \mu_3 \mu_4}$ is non-zero only if all the indices
are equal.

In the form (\ref{eq:diversi}) this equation is not suited for the further
algebraic manipulations that Schoonschip will have to perform. To overcome
this problem we have simply to implement algebraically the various conditions
$\mu_i \neq \mu_j$ in eq.~(\ref{eq:diversi}) by writing
\be
\sin^2 k_{\mu_1} \sin^2 k_{\mu_2} \Big|_{\mu_1 \neq \mu_2} =
\sin^2 k_{\mu_1} \sin^2 k_{\mu_2} -
\sin^4 k_{\mu_1} \: \delta_{\mu_1 \mu_2}
\label{eq:passaggio}
\ee
and the like. This leads to the formula
\bd
{\cal I}(\mu_1,\mu_2,\mu_3,\mu_4) =
\int\! d^{4} k  f(\cos k, \sum_{\lambda} \sin^2 k_{\lambda})
\sin k_{\mu_1} \sin k_{\mu_2} \sin k_{\mu_3}
\sin k_{\mu_4} =
\ed
\bd
\int\! d^{4} k \,  f(\cos k, \sum_{\lambda} \sin^2 k_{\lambda})
\sin^2 k_{\mu_1} \sin^2 k_{\mu_3}
\: \delta_{\mu_1 \mu_2} \: \delta_{\mu_3 \mu_4}
\ed
\bd
+\int\! d^{4} k \,  f(\cos k, \sum_{\lambda} \sin^2 k_{\lambda})
\sin^2 k_{\mu_1} \sin^2 k_{\mu_2}
\: \delta_{\mu_1 \mu_3} \: \delta_{\mu_2 \mu_4}
\ed
\bd
+\int\! d^{4} k \,  f(\cos k, \sum_{\lambda} \sin^2 k_{\lambda})
\sin^2 k_{\mu_1} \sin^2 k_{\mu_2}
\: \delta_{\mu_1 \mu_4} \: \delta_{\mu_2 \mu_3}
\ed
\be
- 2 \cdot \int\! d^{4} k\,  f(\cos k, \sum_{\lambda} \sin^2 k_{\lambda})
\sin^4 k_{\mu_1}
\: \delta_{\mu_1 \mu_2 \mu_3 \mu_4} .
\ee

If $n$ is higher, the way to get useful expressions is the same, but
it is necessary to use relations of the kind (\ref{eq:passaggio})
several times. The expansions for $n$ up to 4 are summarized in
Tables C.1 to C.3.
\vspace{0.5 cm}
\begin{center}
\begin{tabular}{|c|c|c|}
\hline Type of term & Number of permutations & Weight of each permutation \\
\hline ``$\delta_{\mu_1\mu_2} \delta_{\mu_3\mu_4}$'' & 3 & 1 \\
\hline ``$\delta_{\mu_1\mu_2\mu_3\mu_4}$'' & 1 & -2 \\
\hline
\end{tabular}
\end{center}
\begin{center}
{\small {\bf Table C.1} - Numerical coefficients for the expansion of
${\cal I}(\mu_1,\mu_2,\mu_3,\mu_4)$}
\end{center}
\vspace{0.5 cm}
\begin{center}
\begin{tabular}{|c|c|c|}
\hline Type of term & Number of permutations & Weight of each permutation \\
\hline ``$\delta_{\mu_1\mu_2} \delta_{\mu_3\mu_4} \delta_{\mu_5\mu_6}$''
& 15 & 1 \\
\hline ``$\delta_{\mu_1\mu_2} \delta_{\mu_3\mu_4\mu_5\mu_6}$'' & 15 & -2 \\
\hline ``$\delta_{\mu_1\mu_2\mu_3\mu_4\mu_5\mu_6}$'' & 1 & 16 \\
\hline
\end{tabular}
\end{center}
\begin{center}
{\small {\bf Table C.2} - Numerical coefficients for the expansion of
${\cal I}(\mu_1,\mu_2,\mu_3,\mu_4,\mu_5,\mu_6)$}
\end{center}
\vspace{0.5 cm}
\begin{center}
\begin{tabular}{|c|c|c|}
\hline Type of term & Number of permutations & Weight of each permutation \\
\hline ``$\delta_{\mu_1\mu_2} \delta_{\mu_3\mu_4} \delta_{\mu_5\mu_6}
\delta_{\mu_7\mu_8}$'' & 105 & 1 \\
\hline ``$\delta_{\mu_1\mu_2} \delta_{\mu_3\mu_4}
\delta_{\mu_5\mu_6\mu_7\mu_8}$'' & 210 & -2 \\
\hline ``$\delta_{\mu_1\mu_2} \delta_{\mu_3\mu_4\mu_5\mu_6\mu_7\mu_8}$''
& 28 & 16 \\
\hline ``$\delta_{\mu_1\mu_2\mu_3\mu_4} \delta_{\mu_5\mu_6\mu_7\mu_8}$''
& 35 & 4 \\
\hline ``$\delta_{\mu_1\mu_2\mu_3\mu_4\mu_5\mu_6\mu_7\mu_8}$'' & 1 & -272 \\
\hline
\end{tabular}
\end{center}
\begin{center}
{\small {\bf Table C.3} - Numerical coefficients for the expansion of ${\cal I}
(\mu_1,\mu_2,\mu_3,\mu_4,\mu_5,\mu_6,\mu_7,\mu_8)$}
\end{center}

A check on the correctness of
these coefficients may be obtained by considering the case
in which in eq.~(\ref{eq:prodotto}) all indices are equal.
In this case there is only one contribution.
This means that in each Table, if we sum the numbers obtained by multiplying
the weight of each permutation by the number of permutations, we must get 1,
as it is immediately checked in all cases.

\section*{Appendix D}

\section*{$\alpha$-integration}

To reduce the computing time in the numerical integration
of the integrals that involve fermion propagators, we have chosen to
perform analytically the integration over the Feynman parameter $\alpha$, and
to leave for the
numerical integration only a four-dimensional expression. To this end we need
to compute integrals of the form
\be
{\mbox{\large F}}_{n m} \bigg( f(k),g(k)\bigg) = \int_{0}^{1} \! d\alpha \:
\frac{\alpha^{n}}{[f(k) + \alpha g(k)]^{m}} .
\ee
In our calculations we explicitly have
\bea
g(k) & = & 4 \sum_{\lambda} \sin^2 \frac{k_{\lambda}}{2}
- \Big[ \sum_{\lambda} \sin^2 k_{\lambda} + 4 r^2 (\sum_{\lambda}
\sin^2 \frac{k_{\lambda}}{2})^2 \Big] \nonumber \\
f(k) & = & \Big[ \sum_{\lambda} \sin^2 k_{\lambda} + 4 r^2 (\sum_{\lambda}
\sin^2 \frac{k_{\lambda}}{2})^2 \Big] .
\eea
The functions ${\mbox{\large F}}_{n m}$ satisfy the recurrence relations
\be
\frac{\partial}{\partial f} {\mbox{\large F}}_{n m} =
- m \cdot {\mbox{\large F}}_{n\ m+1}
\ee
\be
\frac{\partial}{\partial g} {\mbox{\large F}}_{n m} =
- m \cdot {\mbox{\large F}}_{n+1\ m+1}
\ee
which make simpler their computation and allow for a check of the formulae
given below.

The formulae needed in this work are:
\bea
{\mbox{\large F}}_{0 2} & = & \frac{1}{f \cdot (g+f)} \nonumber \\
{\mbox{\large F}}_{1 2} & = & \frac{1}{g} \cdot \left[ \frac{1}{g}
\log \left( 1+\frac{g}{f} \right) -\frac{1}{g+f} \right] \nonumber \\
{\mbox{\large F}}_{2 2} & = & \frac{1}{g^2} \cdot \left[ 1-2 \frac{f}{g}
\log \left( 1+\frac{g}{f} \right) +\frac{f}{g+f} \right] \nonumber \\
{\mbox{\large F}}_{3 2} & = & \frac{1}{g^2}\cdot \left[ \frac{1}{2}
-2 \frac{f}{g} +3 \frac{f^2}{g^2} \log \left( 1+\frac{g}{f} \right) -
\frac{f^2}{g\cdot (g+f)} \right] \nonumber \\
{\mbox{\large F}}_{0 3} & = & \frac{g+2 f}{2 f^2 \cdot (g+f)^2} \nonumber \\
{\mbox{\large F}}_{1 3} & = & \frac{1}{2 f\cdot (g+f)^2} \nonumber \\
{\mbox{\large F}}_{2 3} & = & \frac{1}{g^2}\cdot \left[ \frac{1}{g}
\log \left( 1+\frac{g}{f} \right) -\frac{3 g+2 f}{2\cdot (g+f)^2} \right]
\nonumber \\
{\mbox{\large F}}_{3 3} & = & \frac{1}{g^3} \cdot \left[ 1-3 \frac{f}{g}
\log \left( 1+\frac{g}{f} \right) +f \frac{5 g+4 f}{2\cdot (g+f)^2} \right]
\nonumber \\
{\mbox{\large F}}_{4 3} & = & \frac{1}{g^3} \cdot \left[ \frac{1}{2}
- 3 \frac{f}{g} + 6 \frac{f^2}{g^2} \log \left( 1+\frac{g}{f} \right)
-f^2 \frac{7 g+6 f}{2 g \cdot (g+f)^2} \right] \nonumber \\
{\mbox{\large F}}_{5 3} & = & \frac{1}{g^3} \cdot \left[ \frac{1}{3}
- \frac{3}{2} \frac{f}{g} + 6 \frac{f^2}{g^2}
- 10 \frac{f^3}{g^3} \log \left( 1+\frac{g}{f} \right)
+f^3 \frac{9 g+8 f}{2 g^2 \cdot (g+f)^2} \right] \nonumber \\
{\mbox{\large F}}_{0 4} & = & \frac{g^2+3 g f+3 f^2}{3 f^3 \cdot (g+f)^3}
\nonumber \\
{\mbox{\large F}}_{1 4} & = & \frac{g+3 f}{6 f^2 \cdot (g+f)^3} \\
{\mbox{\large F}}_{2 4} & = & \frac{1}{3 f\cdot (g+f)^3} \nonumber \\
{\mbox{\large F}}_{3 4} & = & \frac{1}{g^3} \cdot \left[ \frac{1}{g}
\log \left( 1+\frac{g}{f} \right) -\frac{11 g^2+15 g f+6 f^2}{6\cdot (g+f)^3}
\right] \nonumber .
\eea

\end{document}